\documentclass[twocolumn]{aa}  

\usepackage[colorlinks=true,linkcolor=blue,citecolor=blue,urlcolor=blue]{hyperref}

\usepackage{graphicx,natbib,hyperref,epsfig,longtable,txfonts,orcidlink,lineno}
\bibpunct{(}{)}{;}{a}{}{,}

\begin{document}

   \title{Fine-scale downflows above flare ribbons captured by Solar Orbiter/EUI}

   \author{Zheng Sun\inst{1,2,3}\orcidlink{0000-0001-5657-7587}
          \and
          Alexander G.M. Pietrow\inst{3}\orcidlink{0000-0001-5657-7587}
          \and
          Malcolm K. Druett\inst{4}\orcidlink{0000-0001-9845-266X}
          \and
          Hui Tian\inst{1,2}\orcidlink{0000-0002-1369-1758}\thanks{Corresponding author: huitian@pku.edu.cn}
          \and
          Julián D. Alvarado-Gómez\inst{3}\orcidlink{0000-0001-5052-3473}
          \and
          Song Tan\inst{3,5}\orcidlink{0000-0003-0317-0534}
          \and 
          Alexander Warmuth\inst{3}\orcidlink{0000-0003-1439-3610}
          \and 
          Jiasheng Wang\inst{6}\orcidlink{0000-0001-5099-8209}
          \and 
          Yuhang Gao \inst{1,7}\orcidlink{0000-0002-6641-8034}
          \and 
          Zhenyong Hou\inst{1}\orcidlink{0000-0003-4804-5673}
          \and
          Alexandros Stork\inst{3,5}\orcidlink{0009-0009-7147-4258}
          }

   \institute{School of Earth and Space Sciences, Peking University, Beijing 100871, People's Republic of China
         \and
         State Key Laboratory of Solar Activity and Space Weather, National Space Science Center, Chinese Academy of Sciences, Beijing 100190, People's Republic of China
         \and
         Leibniz Institute for Astrophysics Potsdam, An der Sternwarte 16, Potsdam 14482, Germany
         \and
         Plasma Dynamics Group, School of Electrical and Electronic Engineering, University of Sheffield, Sheffield, S1 3JD, UK
         \and 
         Institute of Physics and Astronomy, University of Potsdam, Karl-Liebknecht-Str. 24/25, D-14476 Potsdam-Golm, Germany
         \and
         State Key Laboratory of Solar Activity and Space Weather, National Astronomical Observatories, Chinese Academy of Sciences, Beijing 100101, People's Republic of China
         \and
         Centre for Mathematical Plasma Astrophysics, KU Leuven, Celestijnenlaan 200B, B-3001 Leuven, Belgium
             }

  \abstract
  {In solar flares, flare ribbons map chromospheric footpoints where flare energy deposition occurs. These locations are associated with field aligned energy transport from the corona that results from energy liberated during magnetic reconnection. Recent chromospheric observations in the H$\alpha$ and H$\beta$ bands have revealed fine-scale downflow structures above flare ribbons, referred to as riblets. In this study, we identify similar downflow structures in the extreme-ultraviolet (EUV) wavelength using high-resolution observations from Solar Orbiter/EUI. These fine-scale downflows appear as downward-propagating, bright, and thread-like structures. They exhibit typical velocities of $\sim100~\mathrm{km\ s^{-1}}$, lifetimes of $\sim15$~s, and lengths of $\sim1.6$~Mm. Based on their morphological and dynamical properties, we interpret these observed downflows as the EUV counterparts of the riblets that have previously been reported from chromospheric observations. This study presents EUV imaging of $\sim 10^6$~K downflows above flare ribbons. We interpret these downflows as a result of (1) the energisation and subsequent compression of pre-existing chromospheric fibrils due to particle beams or (2) adiabatic or shock-driven compression induced by the downward-propagating plasma from the corona. These fine-scale EUV riblets provide a new diagnostic tool for probing the dynamics of magnetic reconnection as well as energy transport and deposition during solar flares.}


   \maketitle
%
\graphicspath{{./}}
\section{Introduction} \label{sec:intro}
Solar flares are magnetically driven, impulsive events that can release up to \(10^{32}\) erg and radiate from radio to gamma rays \citep{2011SSRv..159...19F,2017LRSP...14....2B,2025E&PP....9..171L}. Despite extensive observations, the details of energy conversion from magnetic free energy to heating, particle acceleration, and mass motions are not yet fully understood. In the 2D CSHKP framework, the rising of a flux rope forms a current sheet beneath itself; reconnection proceeds within that sheet, and the released energy travels along field lines to the lower atmosphere where it is deposited \citep{1964NASSP..50..451C,1966Natur.211..695S,1974SoPh...34..323H,1976SoPh...50...85K}. Modern 3D studies extend this picture by emphasizing the roles of flux rope topology, quasi-separatrix layers (QSLs), and slipping reconnection in the energy release processes \citep{1996A&A...308..643D,1999A&A...351..707T,2007ApJ...670.1453I,2012A&A...543A.110A,2014ApJ...791L..13L,2015ApJ...804L...8L,2023ApJ...953..148S,2024ApJ...966...70X,2024ScChE..67.1592L,2025ApJ...986...37X,2025A&A...702A..88T}.

The principal agents for field-aligned energy transport in flares are still hotly debated. Beams of energetic particles \citep{2011SSRv..159..107H,2011ApJ...739...96K,2015ApJ...813..113B,2017NatCo...815905D} and thermal conduction \citep{1988ApJ...329..456Z,2012ApJ...754...54B,2021ApJ...912...25A,2026NatAs.tmp....6K} are the most widely studied mechanisms, with some studies also investigating the potential of Alfvénic waves to deliver a non-negligible fraction of the energy \citep{2008Fletcher, 2013Russell, 2016Reep}. Transport of energy from the reconnection site via a combination of beams of energetic particles and thermal conduction has been proposed to deposit energy in the lower atmosphere when encountering higher-density structures. This energy deposition ionises, excites, and heats the local plasma, and also transfers momentum to the plasma within the flux tube. The beam-particle model has been supported by observations of strong hard X-ray (HXR) bremsstrahlung footpoints observed in flares \citep{1981Hoyng, 2002SoPh..210..307F}. The brightening of the lower atmospheric emission at these sites leads to the appearance of elongated structures called flare ribbons. 

\begin{figure*}
   \centering
     \includegraphics[width=0.99\textwidth]{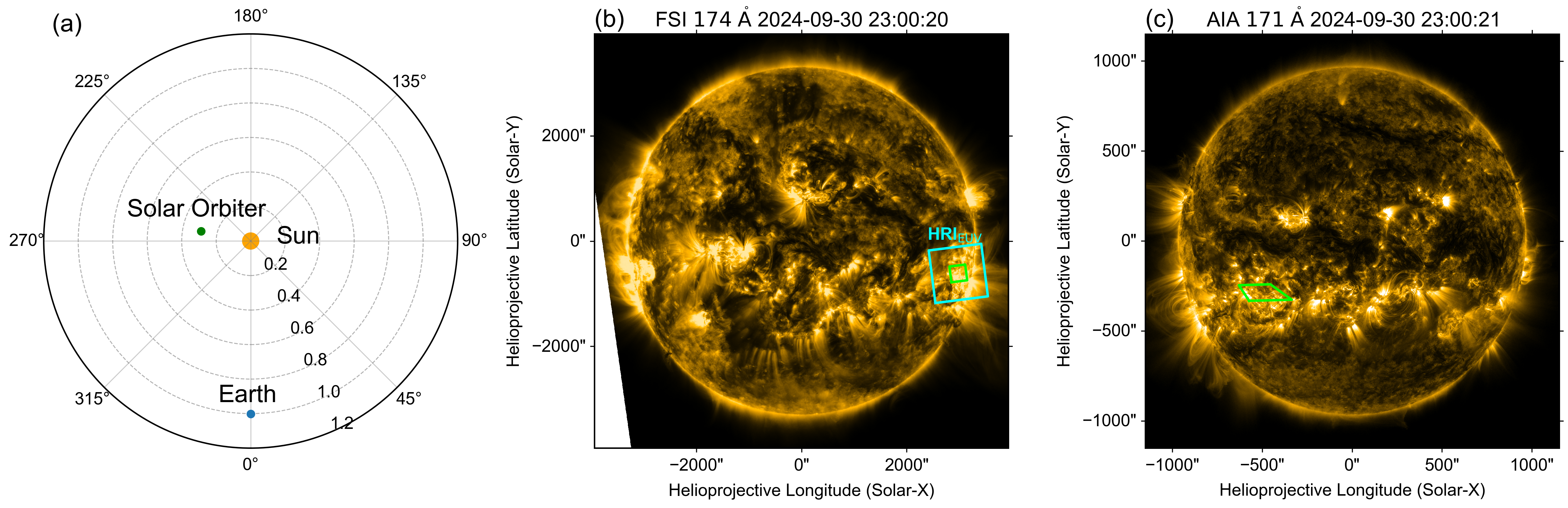}
     \caption{Overview of the positions (a) and viewing geometry from Solar Orbiter (b), and SDO (c). The cyan square denotes the field of view (FoV) of HRI$_{\mathrm{EUV}}$. Since the HRI$_{\mathrm{EUV}}$ FoV partially lies outside the solar disk, we define a smaller on-disk region (green square) within the FoV and re-project this region onto the SDO perspective.}
   \label{fig:Fig.overview}
 \end{figure*}

Ribbons form in the chromosphere on opposite sides of the polarity inversion line (PIL), and subsequently elongate and separate \citep{2002SoPh..210..307F,2011SSRv..159...19F}. Many observations and 3D eruption models show that ribbons frequently exhibit a double-J morphology \citep{1996A&A...308..643D,1999A&A...351..707T,2007ApJ...660..863T,2009ApJ...700..559M,2012A&A...543A.110A,2016ApJ...823...62Z,2017JPlPh..83a5301J,2025ApJ...990...45S}. Observations of emission from ribbons have been interpreted in terms of energy deposition via thermal conduction and beams of particles in 1D models \citep{1973Brown, 1978Emslie, 1981Somov, 1983Ricchiazzi, 1985Fisher, 1987Canfield, 2017NatCo...815905D, 2020Allred} and multi-dimensional models \citep{2020Ruan, 2023Druett, 2024Druett}. One critical feature that tests, and also predicts, behaviors via the modeled energy deposition methods is the generation of bi-directional field-aligned flows known as chromospheric evaporation (upflows) and condensation (downflows). Importantly, the characteristic plasma temperatures of observations that show upflows and downflows during ribbon formation and evolutions have supported a general picture of chromospheric and cool transition region lines ($\log{(T/K)}\approx4.0-5.0$) with downflows \citep{2009Milligan, 2017NatCo...815905D, 2022Kerr, 2022Li, 2023Polito}, and hotter coronal lines ($\log{(T/K)}\approx6.3-7.2$) with upflows \citep{1988ApJ...329..456Z, 2009Milligan,2014ApJ...797L..14T,2015ApJ...811..139T,2019ApJ...879...30L,2022Li,2023ApJ...954....7L}. Lines in the $\log{(T/K)}\approx6.0$ range have typically been reported to show redshifts, e.g. O\,\textsc{vi}--Fe\,\textsc{xiii} lines \citep{2009Milligan}, but little imaging evidence exists to corroborate this.

Recently, high-resolution observations have offered the capability to resolve the fine-scale structure of flare ribbons. Previous studies have reported compact, rapidly evolving kernels or blobs that drift along flare ribbons (e.g., \citealt{2013ApJ...766..127Y,2014ApJ...788L..18S,2022FrASS...940945L,2025A&A...693A...8T,2025ApJ...982L...9Z,2025ApJ...995L..54F}). In addition, \citet{2024A&A...685A.137P} used the Swedish 1-m Solar Telescope \citep[SST; ][]{Scharmer03} to examine the spectral characteristics of substructures within flare ribbons. The effect of these substructures on the average flare spectrum was later followed up in both observations \citep{Pietrow2024,DeWilde2025} and simulations \citep{Yu2025}. Several studies have identified fine-scale upflow or downflow features located above flare ribbons through imaging observations. Using observations from the Interface Region Imaging Spectrograph \citep[IRIS, ][]{2014DePontieuIRIS} and the Atmospheric Imaging Assembly \citep[AIA, ][]{Lemmen2012} on board the Solar Dynamics Observatory \citep[SDO, ][]{Pesnell2012}, \citet{2019PASJ...71...14L} detected jet-like upflows with average velocities of approximately 70~km\,s$^{-1}$, interpreted as signatures of chromospheric evaporation. \citet{2025arXiv250701169S} analyzed high-resolution H$\alpha$ observations from the SST and reported ubiquitous ribbon-localized downflows during an X1.5 flare, with speeds of several tens of km\,s$^{-1}$. They termed these features “riblets” and suggested that they correspond to the redshifted structures observed at the leading edge of flare ribbons \citep{2017NatCo...815905D,2024A&A...685A.137P}. \citet{2026A&A...705A.174T} further conducted a statistical study of riblets using SST observations and identified redshifted emission components with line-of-sight velocities of 16-21~km\,s$^{-1}$ and vertical extents of $620$–$1220~\mathrm{km}$ in H$\beta$. While \citet{2026A&A...705A.113C} recently presented high-cadence, high-resolution imaging from Solar Orbiter showing similar fine-scale downflowing structures in the EUV regime, a detailed quantitative analysis of these features remains absent.
In this paper, we analyze the same eruptive event as \citet{2026A&A...705A.113C} to further investigate these EUV downflows. By utilizing data from Solar Orbiter, we aim to fill the imaging gap in the existing literature regarding ribbon formation dynamics at the $\sim 10^6$~K level. Furthermore, we inspect the physical connectivity between ribbon substructures at chromospheric temperatures ($\log(T/\mathrm{K}) \approx 4.0$) and those at coronal temperatures ($\log(T/\mathrm{K}) \geq 6.0$).

 \begin{figure*}
   \centering
     \includegraphics[width=0.85\textwidth]{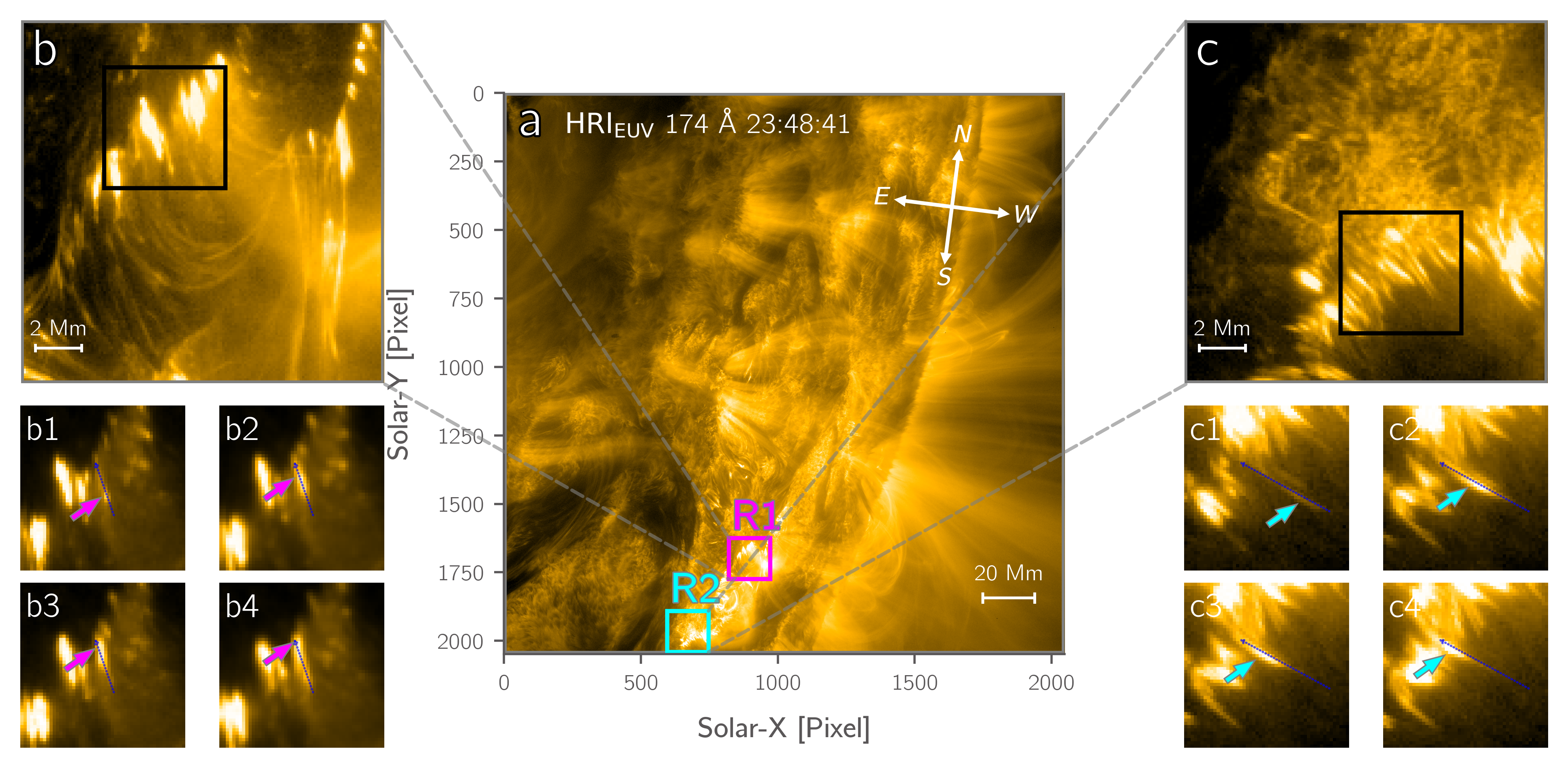}
     \caption{Overview of the downflows in this event. Panel (a) shows the field of view of the HRI$_{\mathrm{EUV}}$ observation. The boxes labeled R1 and R2 mark two different flare-ribbon regions. Panels (b) and (c) present zoomed-in views of regions R1 and R2. Panels (b1–b4) and (c1–c4) display the detailed temporal evolution of two selected individual downflows. The fields of view of these panels are indicated by the black squares in panels (b) and (c), respectively. The blue arrows indicate the trajectories of the downflows. The pink and cyan arrows mark the locations of the downflows in the corresponding frame.} (An online animation is available for this figure.)
   \label{fig:Fig.1}
 \end{figure*}

\section{Observations} 

\label{sec:observation}
On 2024 September 30, an eruptive event occurred in NOAA AR 13842 near the west limb from the perspective of Solar Orbiter, and on the east limb as seen from Earth (See Fig.~\ref{fig:Fig.overview}). 
This event was captured by the High Resolution Imager in the EUV (HRI$_{\mathrm{EUV}}$) telescope of the Extreme Ultraviolet Imager (EUI; \citealt{2020A&A...642A...8R}) instrument onboard the Solar Orbiter. During this time, the spacecraft was located 0.29 au from the Sun, corresponding to a spatial sampling of approximately 0.105 Mm per pixel, a factor of four above the typical AIA resolution of 0.43 Mm per pixel.

HRI$_{\mathrm{EUV}}$ recorded a continuous sequence of images in 174 \AA\, which is most sensitive to plasma around log(T/K) $\approx$ 6.0 \citep{2021A&A...656L...4B,2021A&A...656L...7C}. The data set covers nearly one hour of observations, from 22:55 to 23:55 UT, with a cadence of 2 s and an exposure time of 1.65 s. This exceptional combination of high spatial and temporal resolution enables a detailed inspection of the fine-scale structures of the flare ribbon.
The flux rope began to rise around 23:20 UT and underwent a rapid eruption that continued until approximately 23:55 UT. During the main phase of this activity, an M7.6 flare developed within active region NOAA 13842. The mechanism of energy release associated with this event has been investigated by \citet{2026A&A...705A.113C}.
During the eruption, the filament exhibited a clear untwisting motion and generated a number of small-scale coronal jets, which have been analyzed in recent studies \citep{2025ApJ...985L..12G,2025ApJ...988L..65B,2025A&A...702A.189T,2025MNRAS.544.1758W}. We also employ full-disk 174 Å imaging from the EUI/Full Sun Imager (FSI) onboard Solar Orbiter to provide a global context of the event. A comparison between the full-disk observations from Solar Orbiter and SDO is shown in Figures~\ref{fig:Fig.overview}c and~\ref{fig:Fig.overview}d. To investigate the flare evolution, we also use soft X-ray (SXR) flux observations from the GOES/XRS instrument and HXR observations from the Spectrometer/Telescope for Imaging X-rays \citep[STIX;][]{2020A&A...642A..15K} onboard Solar Orbiter. The GOES/XRS data provide the temporal evolution of the flare emission, while STIX observations are used to identify the timing of the impulsive energy release.

  \begin{figure*}
   \centering
     \includegraphics[width=0.99\textwidth]{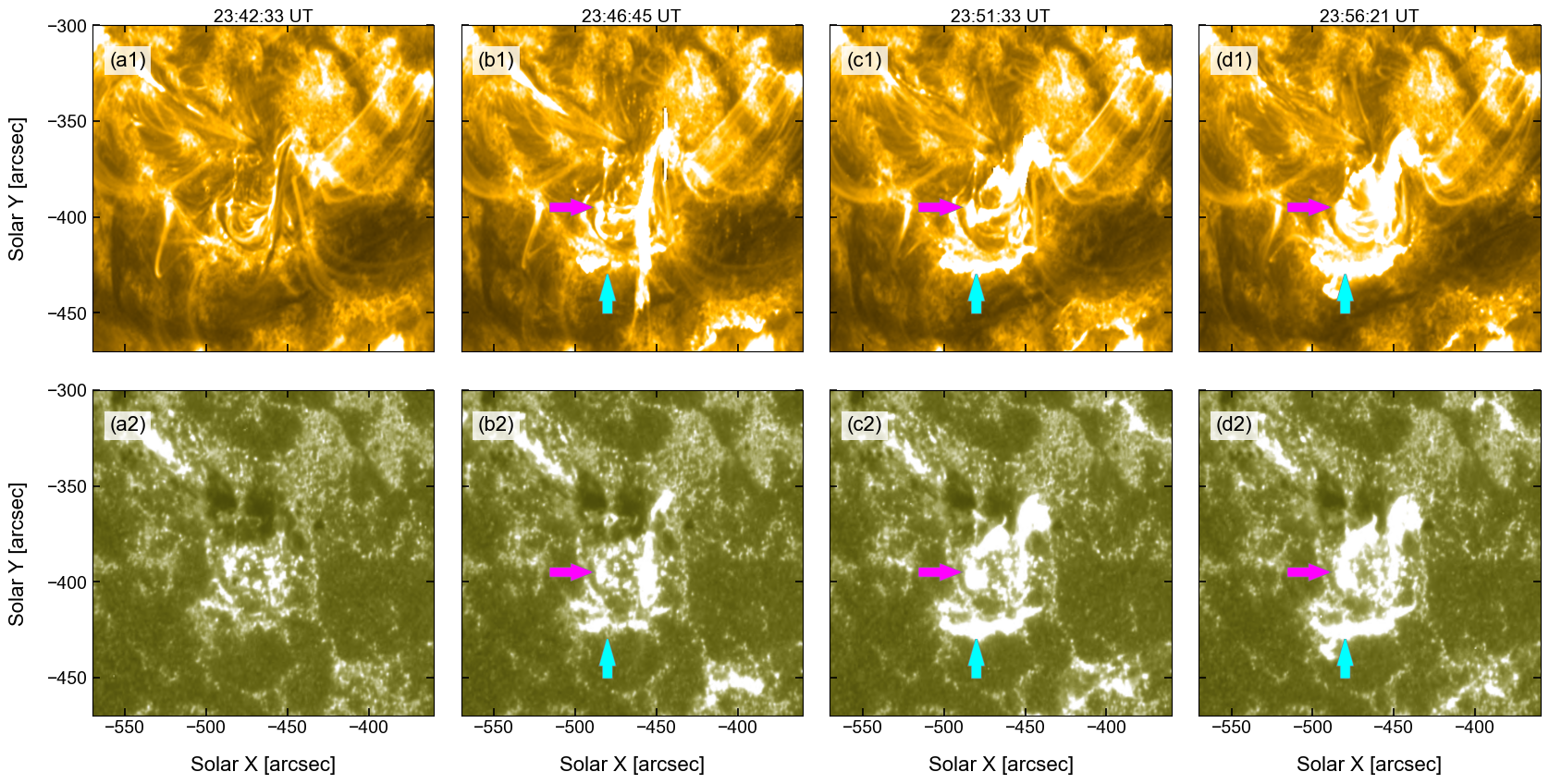}
     \caption{Overview of the same active region} as in Fig.~\ref{fig:Fig.1}, observed with SDO/AIA 171~\AA\ and 1600~\AA\ at four different times. The pink and cyan arrows mark the same locations as in Fig.~\ref{fig:Fig.1}.
   \label{fig:Fig.aia}
 \end{figure*} 

\begin{figure*}
   \centering
     \includegraphics[width=\textwidth]{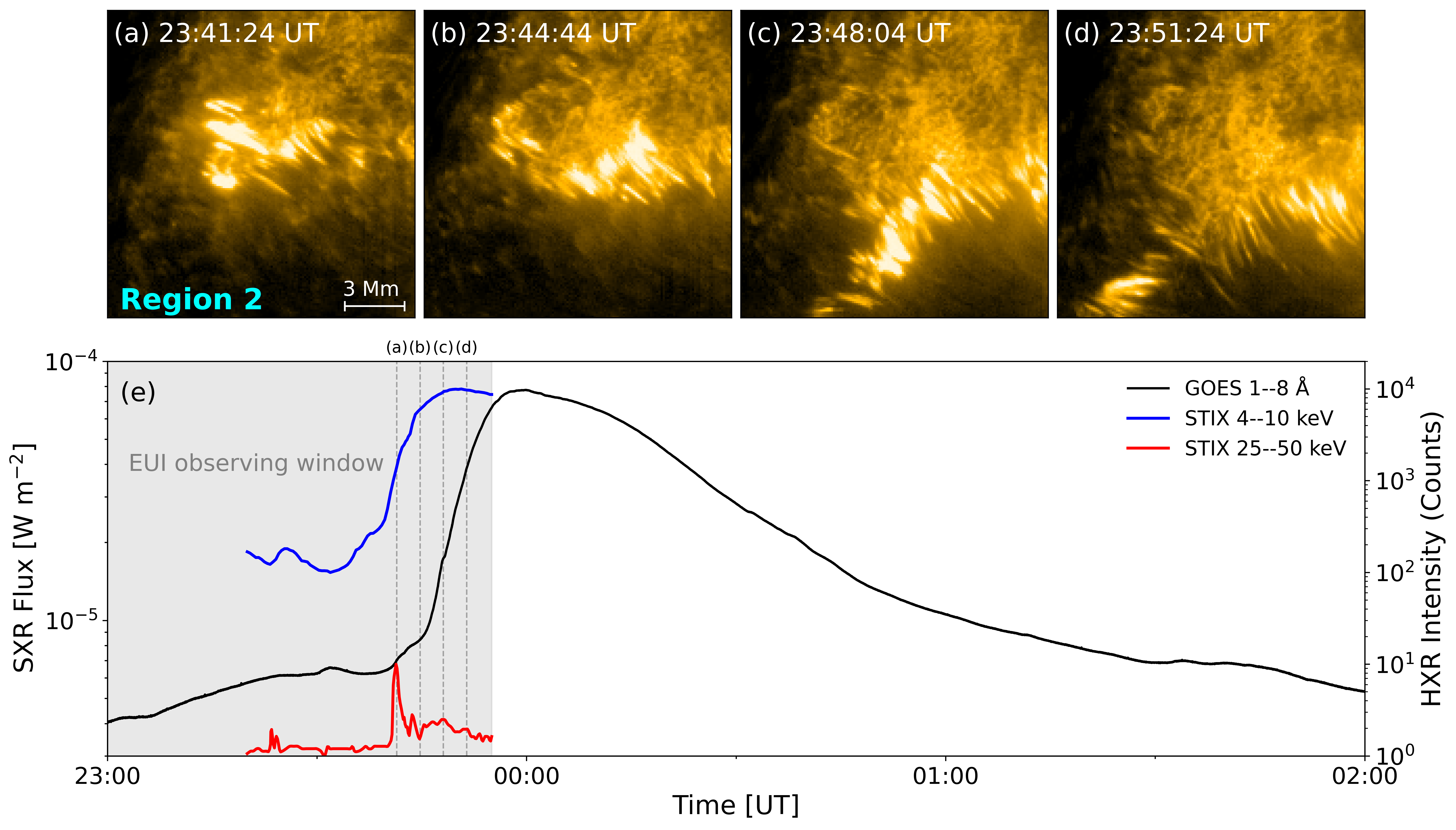}
     \caption{Temporal evolution of Region 2 observed by EUI and the SXR and HXR emission of the flare. Panels (a)–(d) show the evolution of Region 2, while panel (e) displays the GOES 1–8 Å SXR light curve together with the STIX 4–10 keV and 25–50 keV HXR light curves. Details of the HXR data processing are described in \citet{2025A&A...702A.189T}. The gray shaded region indicates the EUI observing time span. The vertical dashed lines mark the times corresponding to panels (a)–(d).
   \label{fig:Fig.2}}
 \end{figure*}

\section{Results} \label{sec:result}
As the filament rises, the flare ribbons appear and develop into several segments of flare ribbons. Fine-scale recurrent downflows above the flare ribbons become discernible starting from approximately 23:38~UT, as also reported by \citet{2026A&A...705A.113C}. Figure~\ref{fig:Fig.1} provides an overview of these downflows. We focus our analysis on two specific areas, Region 1 (R1) and Region 2 (R2), as indicated in the figure. Panels (b1–b4) and (c1-c4) show zoomed-in views of the evolution of two individual downflows in Region 1 and Region 2. These downflows appear as thread-like, slender structures that are bright in $174~\text{\AA}$ and move  toward the flare ribbon. Their morphology and occurrence above the ribbons closely resemble the riblet structures reported by \citet{2025arXiv250701169S} and \citet{2026A&A...705A.174T}. Figure~\ref{fig:Fig.aia} further shows the corresponding locations of the downflows in SDO/AIA $171~\text{\AA}$ and $1600~\text{\AA}$. The ribbons corresponding to regions R1 and R2 are indicated by the pink and cyan arrows, respectively. The 1600 Å images show that the downflows occur in two spatially separated ribbon regions, denoted as R1 and R2. Moreover, it should be noted that the downflows in this event are difficult to resolve with the spatial resolution of AIA. However, some faint brightenings may be present in the unsaturated portions of the flare ribbon, although the available AIA observations do not allow a reliable identification of their nature.

Figures~\ref{fig:Fig.2}(a–d) show the temporal evolution of Region 2. The downflow locations are not stationary, but instead exhibit a gradual displacement over time. This behavior may correspond to successive reconnection between the erupting flux rope and progressively higher overlying magnetic field lines. In panel (e), we plot the temporal evolution of the SXR and HXR emission of the flare. From the figure, we can see that the downflows observed by Solar Orbiter occur during the impulsive phase of the flare.

\begin{figure}
   \centering
     \includegraphics[width=1\columnwidth]{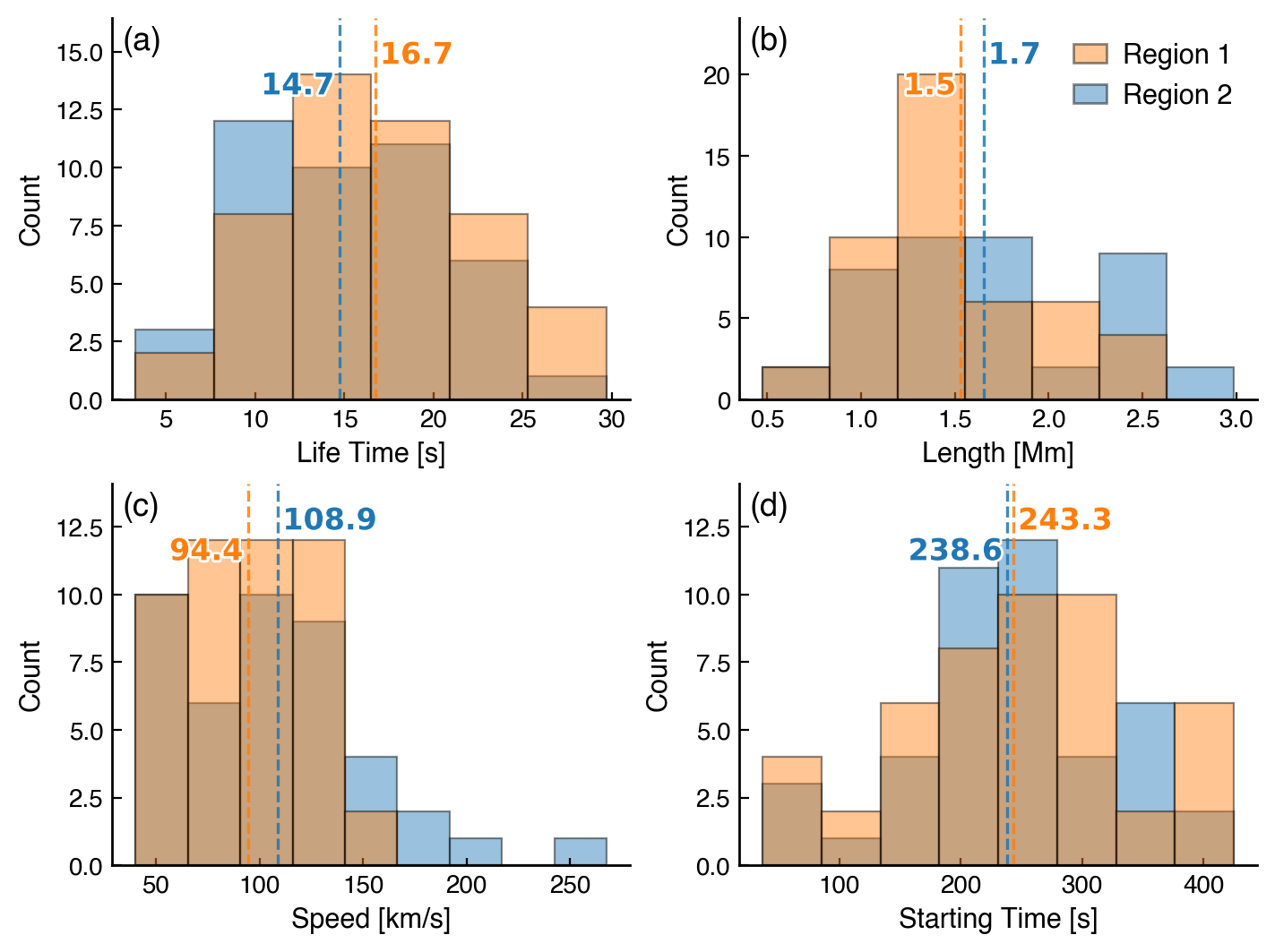}
     \caption{Histograms of the statistical properties of the downflows identified in Region 1 and Region 2. The values marked in each panel indicate the corresponding mean values. The starting time in panel (d) corresponds to the first appearance of the downflow.
   \label{fig:Fig.3}}
 \end{figure}

We note that the riblet-like features in the HRI$_{\mathrm{EUV}}$ images evolve rapidly, making it inaccurate to extract the physical parameters of the downflows by placing fixed slices along their trajectories. To account for this rapid evolution, we instead analyze each downflow manually on a frame-by-frame basis using the open-source software \texttt{Tracker}\footnote{https://opensourcephysics.github.io/tracker-website/}. This approach allows us to determine the lifetime, characteristic width, and projected velocity of individual downflowing features more accurately. An example of the tracking procedure and parameter measurements is presented in Appendix \ref{sec:tracker}. In total, we identified 43 unambiguous downflows in Region~1 and 48 in Region~2. The statistical distributions of their properties are shown in Fig.~\ref{fig:Fig.3}. Since the results from the two regions are basically consistent, we infer that the downflows likely share the same physical origin.
Compared with  \citet{2025arXiv250701169S}, who adopted a similar measurement strategy, the average lifetimes of the downflows in our event are $14.7$ and $16.7$~s for Regions~1 and~2, respectively—close to their reported value of $\sim12$~s. Such short timescales indicate that these features are difficult to be detected by SDO/AIA, which has a cadence of 12~s.   
The lengths of the downflows are $1.5$ and $1.7$~Mm, slightly larger than the $\sim0.95$~Mm reported by \citet{2025arXiv250701169S}$,$ and $0.62$–$1.2$~Mm measured by \citet{2026A&A...705A.174T}. The averaged velocities, $91$ and $100~\mathrm{km\ s^{-1}}$ for the two regions, exceed the $\sim47~\mathrm{km\ s^{-1}}$ reported in \citet{2025arXiv250701169S}. The starting times of the downflows are distributed almost uniformly throughout the entire eruption. 

Overall, the observed lifetimes, lengths, and velocities resemble the properties of the riblets identified in chromospheric lines by \citet{2025arXiv250701169S} and \citet{2026A&A...705A.174T}. We therefore interpret the fine-scale downflows reported here as the EUV counterparts of riblets.

\section{Discussion} \label{sec:discussion}
In this section, we discuss our observations and outline several possible physical origins of these manifestations, together with suggestions for future observations that could help discriminate between these scenarios.

\subsection{EUV ribbon substructures and the association with chromospheric riblets}

This set of observations extends the observational coverage of flare-related downflows from previously reported condensation temperatures at $\sim 10^5$~K to and evaporation at coronal temperatures of $\sim 1 \times 10^6$~K for flare ribbon formation such as \cite{2009Milligan}, \cite{2023Polito}, and \cite{2024A&A...685A.137P}. We observe widespread downflowing  plasma in the $\sim 1 \times 10^6$~K range at the footpoints of flare loops. This information is critical to test, validate and discriminate among energy transport, deposition, and atmospheric response theories via simulations. 

Additionally, thanks to recent sub-second chromospheric imaging of riblets we are able to compare the dynamics of $\sim 1 \times 10^6$~K plasma with that of $\sim 1 \times 10^4$~K plasma \citep{2025arXiv250701169S,2026A&A...705A.174T}. The structures we report have sufficiently similar morphology, lifetimes, and locations to infer that these features are likely EUV riblets, that are higher, hotter extensions of chromospheric riblet fine structures. It should also be noted that the 174~\AA\ passband has a broad temperature response extending well beyond its peak sensitivity near $\sim1$~MK. Therefore, the structures identified here may correspond to the same physical features previously observed in cooler chromospheric lines, viewed through a different temperature response regime. Differences in reported lengths and velocity may arise from (1) the gradients in the evolving plasma, where the physically higher, longer, and hotter components have greater velocities than the lower, cooler, regions which have stronger pressure gradients or (2) from different projection geometries between these events. Again, this issue could be more conclusively addressed through coordinated multi-wavelength observations covering a broader temperature range, combining Solar Orbiter with chromosphere-targeting telescopes and future spectroscopic diagnostics (e.g., MUSE, EUVST).

\subsection{Physical interpretation of the riblets}
Such connected fine structures imply that flare energy deposition drives coherent plasma structures propagating downwards during ribbon formation that contains multi-thermal plasma with temperatures between $\sim 1 \times 10^4$~K and $\sim 1 \times 10^6$~K. Previous chromospheric riblet studies have attempted to explain the physical origin of riblets: \citet{2025arXiv250701169S} suggested that the riblets correspond to redshifted destruction of chromospheric structures at the expanding edges of flare ribbons \citep{2017NatCo...815905D,2024A&A...685A.137P}, whereas \citet{2026A&A...705A.174T} proposed that riblets are associated with tearing-mode instabilities in the reconnection current sheet \citep{2025ApJ...993...31D} to account for the recurrent nature of the downflows.
However, these interpretations primarily focus on the association with magnetic reconnection but do not explicitly address the mechanisms by which flare energy is transported to the lower atmosphere to produce the observed downflows.
In the following, we propose two scenarios to explain these observations.

\subsubsection{Possible Scenario 1: energisation and compression of fibrils}

If flare ribbon formation compresses former active region chromospheric structures to lower geometric heights, then the main candidates showing correspondence with riblets are the fibrils (or spicules when seen from the side). Chromospheric fibrils extend upward from the chromosphere toward the low corona. Dynamic fibrils in active regions have typical lengths of $\sim1250$~km, but can vary with local conditions, ranging from $400$ to $5200$~km when observed in H$\alpha$ \citep{2007DePontieu,2019Sci...366..890S,2026A&A...706A.369J,chen2026spicules} or analysed in magnetohydrodynamic simulations \citep{2006Hansteen,2025A&A...701A.294C}. Under normal conditions they are theorised to be launched due to the actions of p-modes or magnetic reconnection, and have parabolic paths in height-time diagrams with maximum ascent or descent velocities achieved near launch or termination, of $15-20$~km\,s$^{-1}$ but reaching to $35$~km\,s$^{-1}$ in some cases. Fibrils are ubiquitously distributed throughout active regions (see Fig. 7 of \citealt{2024A&A...685A.137P}). The lengths of these structures are consistent with the observed features. However, the observed higher velocities and non-parabolic height-time trajectories suggest that these flows are driven by additional energy deposition and plasma compression from the flare, rather than the simple ballistic collapse of existing structures. Additionally this picture is corroborated by the lower subsequent formation heights of flare ribbon emission \citep{2012MartinezOliveros, 2015Krucker,2016ApJ...823L..13L, 2020Kuridze, 2024A&A...685A.137P} as well as geometric observational arguments from overlapping structures in flare observations \citep{2024A&A...685A.137P}.

\begin{figure}
   \centering
     \includegraphics[width=1\columnwidth]{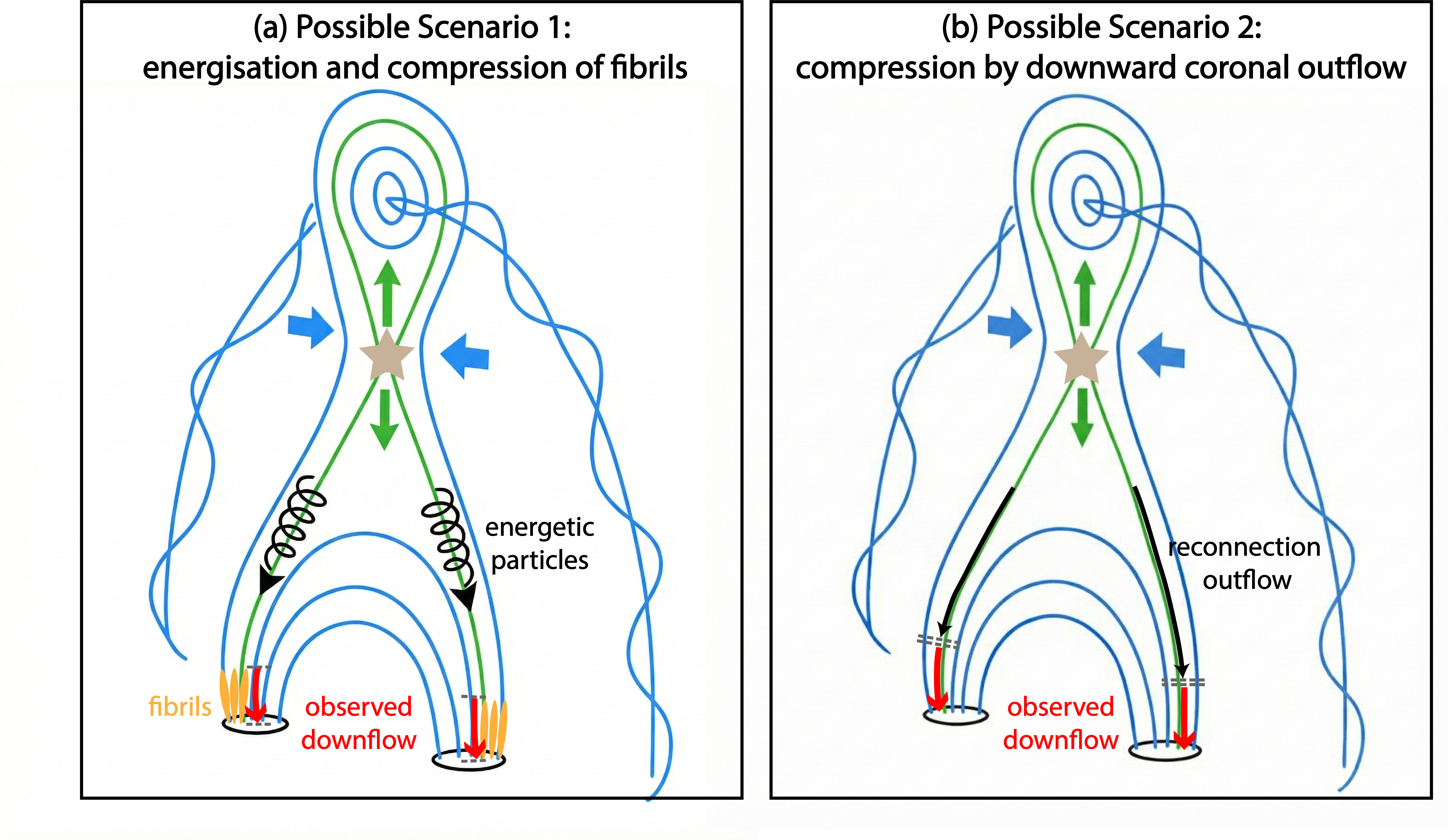}
     \caption{Cartoon scenarios of the observed downflows. The five-pointed star symbol indicates the magnetic reconnection site. This sketch is adapted from \citet{lysenko2020x}.
   \label{fig:Fig.cartoon}}
 \end{figure}

When newly reconnected flare field lines in the current sheet retract and relax, they form coronal loops anchored at their chromospheric footpoints, which correspond to the leading edges of the flare ribbons. This magnetic restructuring significantly alters the surrounding environment of the pre-existing fibrils. Energy is deposited via field-aligned thermal conduction or non-local transport by energetic particle beams from the reconnection site, leading to the ionization, excitation, and heating of the flux-tube plasma, as well as the transfer of momentum. 
Under the thermal conduction scenario, the observed downward motion may not represent a bulk physical plasma flow, but rather the propagation of a heating front \citep{2017ApJ...845L..18D}. In this framework, the downflow is an apparent motion where the transition region is pushed toward lower, denser atmospheric layers. For this interpretation to be viable, the measured downward velocities of the bright features must be consistent with the propagation speed of a conduction front as it heats the plasma to EUV-emitting temperatures.
On the other hand, recent on-disk Doppler shift observations by \citet{2024A&A...685A.137P} and side-view observations by \citet{2025arXiv250701169S,2026A&A...705A.174T} show mutually consistent downflow velocities, suggesting that these features are likely associated with real plasma motions. Given the strong morphological similarities between those chromospheric riblets and the structures reported here, our observations may also favor a physical downflow interpretation. Nevertheless, we cannot fully exclude that some of the observed motions are related to the propagation of heating fronts. A schematic illustration of the beam-driven scenario is presented in Figure~\ref{fig:Fig.cartoon}a.

\subsubsection{Possible Scenario 2: compression by downward coronal outflow}
When magentic reconnection occurs high in the corona, reconnection outflows are generated \citep{1976SoPh...50...85K,2000mare.book.....P}. These outflows may then propagate downward along the newly formed post flare loops into the denser lower atmosphere (e.g., \citealt{2014ApJ...797L..14T}). During this process, they can undergo adiabatic compression, which increases the plasma density and temperature and may lead to the formation of bright downflowing structures within the flux tube. Another related possibility is that the enhanced density in the low atmosphere reduces the local Alfvén speed relative to that in the corona \citep{1981SoPh...70...25H,2011ApJ...736....3V}. In this case, the downward outflows may exceed the local Alfvén speed near the chromosphere and form a termination shock \citep{2001EP&S...53..473S,2002A&A...384..273A,2009A&A...494..669M,2019MNRAS.489.3183C}. Such a shock would compress the ambient plasma and increase its temperature and density \citep{1983SoPh...84..169F,2015ApJ...805..135T}, allowing the downflows to appear as bright EUV structures in low altitudes. In addition, similar to Scenario 1, the observed downward motion may not necessarily correspond entirely to bulk plasma flows. Part of the apparent motion could also be associated with the propagation of a heating or compression front along the magnetic structure, which may contribute to the observed EUV brightening. The cartoon illustration of this scenario is shown in Figure~\ref{fig:Fig.cartoon}b.

An alternative possible origin of the observed downflows is coronal rain, which also consists of downflowing plasma along post-flare loops, typically occurring during the decay phase of flares \citep{2016ApJ...833..184S,2021ApJ...920L..15R,2024ApJ...970..106S}. To examine this possibility, we compare the occurrence times of the downflows with the GOES soft X-ray and STIX hard X-ray light curves (Figure \ref{fig:Fig.2}. We find that the downflows appear almost simultaneously with the flare onset, and a large number of downflows are already present during the impulsive phase. This temporal behavior is inconsistent with the typical evolution of coronal rain, which is generally associated with cooling processes during the later decay phase (e.g. \citealt{2014ApJ...780L..28M,2022A&A...659A.107C,2024ApJ...973...57Q}). Therefore, we consider it unlikely that coronal rain is the dominant mechanism responsible for the downflows analyzed in this work.

The coronal-outflow-driven compression scenario is also supported by large-scale downflow observations \citep{2024ApJ...974..205S,2025ApJ...987..115W}. \citet{2024ApJ...974..205S} observed large-scale downflows near the footpoints of a polar crown filament. The downflows can be observed across all SDO/AIA passbands with speeds of \(92\text{–}144~\mathrm{km\ s^{-1}}\), and similar phenomena were later found in several filament eruptions \citep{2025ApJ...987..115W}. These downflows share several key features with the riblet-like structures reported here: they occur along flare ribbons, manifest as EUV bright structures, appear only at relatively low altitudes, exhibit thread-like EUV morphology, and show recurrent behavior. In those events, the heights of the downflows are far beyond the scale of chromospheric fibrils and therefore cannot be explained by Scenario 1. Instead, they were interpreted as a consequence of the conversion of kinetic energy from falling filament material into thermal energy, supported by the clear observational evidence that the downflows originate directly from the descending filament material (see Fig. 2(g) in \citealt{2024ApJ...974..205S}). By analogy, a similar physical mechanism may operate in our event. However, instead of a massive amount of filament material falling back to the lower atmosphere, the falling structures here would represent downward reconnection outflows. Because reconnection outflows are far less massive than filament plasma, the resulting downflows would naturally be smaller in scale and less pronounced than those associated with filament eruptions.

Additional hints supporting this scenario might also be found in flare numerical simulations (e.g.\ \citealt{2023ApJ...947...67R,2023ApJ...955...88Y,2024Druett}). These models show that reconnection outflows impact the top of the flare arcade and subsequently propagate downward along the arcade. However, whether such downward-propagating flows can further compress plasma at lower heights and manifest as the downflows observed here remains unclear, and will require future numerical simulations to be investigated in detail.

\subsection{Connection to three-dimensional eruption models}

In Fig.~\ref{fig:Fig.2}(a--d) we observe continuous drifts in the locations of the downflows. This behavior may correspond to successive reconnection between the erupting flux rope and progressively higher overlying magnetic field lines, causing the corresponding magnetic footpoints to move farther away from the eruption center. This suggests that such downflows could serve as observational tracers of the evolution of magnetic reconnection during eruptive flares and may help us better understand the different reconnection processes involved in flux rope eruptions.

Additionally, the recurrent nature of the observed downflows may be explained by tearing-mode instabilities developing in the reconnection current sheet, as also suggested by \citet{2026A&A...705A.174T}.
\citet{2025ApJ...993...31D} performed three-dimensional magnetohydrodynamic eruption simulations and found that tearing-mode instabilities arise naturally in turbulent current sheets. The resulting intermittent reconnection can produce recurrent fine structures in flare ribbons. However, their analysis focused primarily on horizontal ribbon substructures, rather than on vertical motions such as the downflows reported here.
Nevertheless, if reconnection energy release is intermittent, it is expected to generate intermittent particle beams and downward reconnection outflows (plasmoids) as well. These intermittent energy and momentum injections could then lead to the recurrent downflows observed in this work. If this interpretation is correct, the waiting times and occurrence frequencies of the downflows may provide valuable observational constraints on tearing-mode-driven magnetic reconnection.

\section{Conclusion}

In this work, we use Solar Orbiter/EUI observations to report fine-scale downflow structures above flare ribbons in EUV 174 \AA. These features appear as downward-moving, bright, thread-like structures with speeds of $\sim$100~km\,s$^{-1}$, lifetimes of $\sim$15~s, and lengths of $\sim$1.6~Mm. We interpret these structures as EUV riblets, the coronal $\sim 10^6$~K fine structure extensions of the chromospheric $\sim 10^4$~K ribbon fine-structures observed in chromospheric lines \citep{2025arXiv250701169S,2026A&A...705A.174T}. Their spatial and temporal scales are extremely small, making such fine structures difficult to be resolved with SDO/AIA. In contrast, the high-resolution EUV observations from Solar Orbiter/EUI allow us to identify these downflows clearly. 

We discuss two possible scenarios for the formation of these structures:
Flare-related energy release, in the form of energetic particle beams (Scenario~1) or downward plasma drains (Scenario~2), can lead to energy deposition at relatively low atmospheric heights, manifesting observationally as the downflows reported here. The observed downward motions may correspond to real plasma flows or to the propagation of heating fronts, and the current observations do not allow us to directly distinguish between these possibilities. Because both scenarios are intrinsically linked to magnetic reconnection, these downflows are expected to occur at reconnection footpoints, that is, along flare ribbons. The recurrent nature of the downflows may further be explained by tearing-mode instabilities operating in the reconnection current sheet.

We suggest that such fine-scale downflows may be a ubiquitous phenomenon accompanying solar flares. Confirming this possibility will require additional high-resolution observations, particularly those capable of resolving the relevant spatial and temporal scales. Systematic campaigns with Solar Orbiter/EUI and co-observations with high-resolution ground-based telescopes capable of observing the chromosphere will be essential to determine how commonly these structures occur and how they relate to the energy release and plasma dynamics in flares. Moreover, $\sim 10^6$~K downflows provide an observational constraint that simulations should test, namely whether plasma in this temperature response channel can produce downflowing substructures when energy is transported primarily via the proposed mechanisms. This helps to improve our understanding of energy transport processes during solar flares.

\begin{acknowledgements}
This work is supported by the Strategic Priority Research Program of the Chinese Academy of Sciences (grant No. XDB0560000), the National Natural Science Foundation of China (12425301, 425B2024), China’s Space Origins Exploration Program, and the Specialized Research Fund for State Key Laboratory of Solar Activity and Space Weather. AP is supported by grant PI~2102/1-1 from the Deutsche Forschungsgemeinschaft (DFG). H.T. is also supported by the New Cornerstone Science Foundation through the Xplorer Prize. 
\end{acknowledgements}

\bibliographystyle{aa} 
\bibliography{ref}

@INPROCEEDINGS{Scharmer03,
       author = {{Scharmer}, Goran B. and {Bjelksjo}, Klas and {Korhonen}, Tapio K. and
         {Lindberg}, Bo and {Petterson}, Bertil},
        title = "{The 1-meter Swedish solar telescope}",
    booktitle = {Innovative Telescopes and Instrumentation for Solar Astrophysics},
         year = "2003",
       editor = {{Keil}, Stephen L. and {Avakyan}, Sergey V.},
       series = {Proc. SPIE},
       volume = {4853},
        month = "Feb",
        pages = {341-350},
          doi = {10.1117/12.460377},
       adsurl = {https://ui.adsabs.harvard.edu/abs/2003SPIE.4853..341S}
}

@ARTICLE{Yu2025,
       author = {{Yu}, H.~C. and {Hong}, J. and {Ding}, M.~D.},
        title = "{Sun-as-a-star analysis of simulated solar flares}",
      journal = {\aap},
         year = 2025,
        month = feb,
       volume = {694},
          eid = {A315},
        pages = {A315},
          doi = {10.1051/0004-6361/202451706},
archivePrefix = {arXiv},
       eprint = {2502.00441},
 primaryClass = {astro-ph.SR},
       adsurl = {https://ui.adsabs.harvard.edu/abs/2025A&A...694A.315Y}
}

@ARTICLE{DeWilde2025,
       author = {{De Wilde}, M. and {Pietrow}, A.~G.~M. and {Druett}, M.~K. and {Pastor Yabar}, A. and {Koza}, J. and {Kontogiannis}, I. and {Andriienko}, O. and {Berlicki}, A. and {Brunvoll}, A.~R. and {de la Cruz Rodr{\'\i}guez}, J. and {Thoen Faber}, J. and {Joshi}, R. and {Kuridze}, D. and {N{\'o}brega-Siverio}, D. and {Rouppe van der Voort}, L.~H.~M. and {Ryb{\'a}k}, J. and {Scullion}, E. and {Silva}, A.~M. and {Vashalomidze}, Z. and {Vicente Ar{\'e}valo}, A. and {Wi{\'s}niewska}, A. and {Yadav}, R. and {Zaqarashvili}, T.~V. and {Zbinden}, J. and {{\O}yre}, E.~S.},
        title = "{Synthesizing Sun-as-a-star flare spectra from high-resolution solar observations}",
      journal = {\aap},
         year = 2025,
        month = aug,
       volume = {700},
          eid = {A275},
        pages = {A275},
          doi = {10.1051/0004-6361/202554870},
archivePrefix = {arXiv},
       eprint = {2507.07967},
 primaryClass = {astro-ph.SR},
       adsurl = {https://ui.adsabs.harvard.edu/abs/2025A&A...700A.275D}
}

@ARTICLE{2024ScChE..67.1592L,
       author = {{Li}, Dong and {Hou}, ZhenYong and {Bai}, XianYong and {Li}, Chuan and {Fang}, Matthew and {Zhao}, HaiSheng and {Wang}, JinCheng and {Ning}, ZongJun},
        title = "{Simultaneous detection of flare-associated kink oscillations and extreme-ultraviolet waves}",
      journal = {Science in China E: Technological Sciences},
         year = 2024,
        month = may,
       volume = {67},
       number = {5},
        pages = {1592-1601},
          doi = {10.1007/s11431-023-2534-8},
archivePrefix = {arXiv},
       eprint = {2311.08767},
 primaryClass = {astro-ph.SR},
       adsurl = {https://ui.adsabs.harvard.edu/abs/2024ScChE..67.1592L}
}

@ARTICLE{2025E&PP....9..171L,
       author = {{Li}, YiYang and {Huang}, ShiYong and {Xu}, SiBo and {Yuan}, ZhiGang and {Jiang}, Kui and {Xiong}, QiYang and {Lin}, RenTong},
        title = "{Solar flare forecasting based on a Fusion Model}",
      journal = {Earth and Planetary Physics},
         year = 2025,
        month = jan,
       volume = {9},
       number = {1},
        pages = {171-181},
          doi = {10.26464/epp2024058},
       adsurl = {https://ui.adsabs.harvard.edu/abs/2025E&PP....9..171L}
}

@ARTICLE{2011SSRv..159...19F,
       author = {{Fletcher}, L. and {Dennis}, B.~R. and {Hudson}, H.~S. and {Krucker}, S. and {Phillips}, K. and {Veronig}, A. and {Battaglia}, M. and {Bone}, L. and {Caspi}, A. and {Chen}, Q. and {Gallagher}, P. and {Grigis}, P.~T. and {Ji}, H. and {Liu}, W. and {Milligan}, R.~O. and {Temmer}, M.},
        title = "{An Observational Overview of Solar Flares}",
      journal = {\ssr},
         year = 2011,
        month = sep,
       volume = {159},
       number = {1-4},
        pages = {19-106},
          doi = {10.1007/s11214-010-9701-8},
archivePrefix = {arXiv},
       eprint = {1109.5932},
 primaryClass = {astro-ph.SR},
       adsurl = {https://ui.adsabs.harvard.edu/abs/2011SSRv..159...19F}
}

@ARTICLE{2017LRSP...14....2B,
       author = {{Benz}, Arnold O.},
        title = "{Flare Observations}",
      journal = {Living Reviews in Solar Physics},
         year = 2017,
        month = dec,
       volume = {14},
       number = {1},
          eid = {2},
        pages = {2},
          doi = {10.1007/s41116-016-0004-3},
       adsurl = {https://ui.adsabs.harvard.edu/abs/2017LRSP...14....2B}
}

@ARTICLE{1976SoPh...50...85K,
       author = {{Kopp}, R.~A. and {Pneuman}, G.~W.},
        title = "{Magnetic reconnection in the corona and the loop prominence phenomenon.}",
      journal = {\solphys},
         year = 1976,
        month = sep,
       volume = {50},
       number = {1},
        pages = {85-98},
          doi = {10.1007/BF00206193},
       adsurl = {https://ui.adsabs.harvard.edu/abs/1976SoPh...50...85K}
}

@ARTICLE{1974SoPh...34..323H,
       author = {{Hirayama}, T.},
        title = "{Theoretical Model of Flares and Prominences. I: Evaporating Flare Model}",
      journal = {\solphys},
         year = 1974,
        month = feb,
       volume = {34},
       number = {2},
        pages = {323-338},
          doi = {10.1007/BF00153671},
       adsurl = {https://ui.adsabs.harvard.edu/abs/1974SoPh...34..323H}
}

@ARTICLE{1966Natur.211..695S,
       author = {{Sturrock}, P.~A.},
        title = "{Model of the High-Energy Phase of Solar Flares}",
      journal = {\nat},
         year = 1966,
        month = aug,
       volume = {211},
       number = {5050},
        pages = {695-697},
          doi = {10.1038/211695a0},
       adsurl = {https://ui.adsabs.harvard.edu/abs/1966Natur.211..695S}
}

@INCOLLECTION{1964NASSP..50..451C,
       author = {{Carmichael}, H.},
        title = "{A Process for Flares}",
    booktitle = {NASA Special Publication},
         year = 1964,
       editor = {{Hess}, Wilmot N.},
       volume = {50},
        pages = {451},
       adsurl = {https://ui.adsabs.harvard.edu/abs/1964NASSP..50..451C}
}

@ARTICLE{2007ApJ...670.1453I,
       author = {{Isenberg}, Philip A. and {Forbes}, Terry G.},
        title = "{A Three-dimensional Line-tied Magnetic Field Model for Solar Eruptions}",
      journal = {\apj},
         year = 2007,
        month = dec,
       volume = {670},
       number = {2},
        pages = {1453-1466},
          doi = {10.1086/522025},
       adsurl = {https://ui.adsabs.harvard.edu/abs/2007ApJ...670.1453I}
}

@ARTICLE{2012A&A...543A.110A,
       author = {{Aulanier}, G. and {Janvier}, M. and {Schmieder}, B.},
        title = "{The standard flare model in three dimensions. I. Strong-to-weak shear transition in post-flare loops}",
      journal = {\aap},
         year = 2012,
        month = jul,
       volume = {543},
          eid = {A110},
        pages = {A110},
          doi = {10.1051/0004-6361/201219311},
       adsurl = {https://ui.adsabs.harvard.edu/abs/2012A&A...543A.110A}
}

@ARTICLE{2025MNRAS.544.1758W,
       author = {{Wallace}, Tarhik and {Antolin}, Patrick},
        title = "{Reconnection nanojets associated with a prominence eruption observed with Solar Orbiter/EUI-HRI}",
      journal = {\mnras},
         year = 2025,
        month = dec,
       volume = {544},
       number = {2},
        pages = {1758-1768},
          doi = {10.1093/mnras/staf1879},
archivePrefix = {arXiv},
       eprint = {2510.25068},
 primaryClass = {astro-ph.SR},
       adsurl = {https://ui.adsabs.harvard.edu/abs/2025MNRAS.544.1758W}
}

@ARTICLE{2025ApJ...988L..65B,
       author = {{Bura}, Annu and {Shrivastav}, Arpit Kumar and {Patel}, Ritesh and {Samanta}, Tanmoy and {Nayak}, Sushree S. and {Ghosh}, Ananya and {Sow Mondal}, Shanwlee and {Pant}, Vaibhav and {Seaton}, Daniel B.},
        title = "{Dynamics of Reconnection Nanojets in Eruptive and Confined Solar Flares}",
      journal = {\apjl},
         year = 2025,
        month = aug,
       volume = {988},
       number = {2},
          eid = {L65},
        pages = {L65},
          doi = {10.3847/2041-8213/adef0f},
archivePrefix = {arXiv},
       eprint = {2507.04639},
 primaryClass = {astro-ph.SR},
       adsurl = {https://ui.adsabs.harvard.edu/abs/2025ApJ...988L..65B}
}

@ARTICLE{2025A&A...702A.189T,
       author = {{Tan}, Song and {Warmuth}, Alexander and {Schuller}, Fr{\'e}d{\'e}ric and {Shen}, Yuandeng and {Mitchell}, Jake A.~J. and {Shi}, Fanpeng},
        title = "{Extremely diverse coronal jets accompanying an erupting filament captured by Solar Orbiter}",
      journal = {\aap},
         year = 2025,
        month = oct,
       volume = {702},
          eid = {A189},
        pages = {A189},
          doi = {10.1051/0004-6361/202555297},
archivePrefix = {arXiv},
       eprint = {2509.04741},
 primaryClass = {astro-ph.SR},
       adsurl = {https://ui.adsabs.harvard.edu/abs/2025A&A...702A.189T}
}

@ARTICLE{2025ApJ...985L..12G,
       author = {{Gao}, Yuhang and {Tian}, Hui and {Berghmans}, David and {Duan}, Yadan and {Van Doorsselaere}, Tom and {Chen}, Hechao and {Kraaikamp}, Emil},
        title = "{Reconnection Nanojets in an Erupting Solar Filament with Unprecedented High Speeds}",
      journal = {\apjl},
         year = 2025,
        month = may,
       volume = {985},
       number = {1},
          eid = {L12},
        pages = {L12},
          doi = {10.3847/2041-8213/add33a},
archivePrefix = {arXiv},
       eprint = {2504.20663},
 primaryClass = {astro-ph.SR},
       adsurl = {https://ui.adsabs.harvard.edu/abs/2025ApJ...985L..12G}
}

@ARTICLE{2021A&A...656L...7C,
       author = {{Chen}, Yajie and {Przybylski}, Damien and {Peter}, Hardi and {Tian}, Hui and {Auch{\`e}re}, F. and {Berghmans}, D.},
        title = "{Transient small-scale brightenings in the quiet solar corona: A model for campfires observed with Solar Orbiter}",
      journal = {\aap},
         year = 2021,
        month = dec,
       volume = {656},
          eid = {L7},
        pages = {L7},
          doi = {10.1051/0004-6361/202140638},
archivePrefix = {arXiv},
       eprint = {2104.10940},
 primaryClass = {astro-ph.SR},
       adsurl = {https://ui.adsabs.harvard.edu/abs/2021A&A...656L...7C}
}

@ARTICLE{2021A&A...656L...4B,
       author = {{Berghmans}, D. and {Auch{\`e}re}, F. and {Long}, D.~M. and {Soubri{\'e}}, E. and {Mierla}, M. and {Zhukov}, A.~N. and {Sch{\"u}hle}, U. and {Antolin}, P. and {Harra}, L. and {Parenti}, S. and {Podladchikova}, O. and {Aznar Cuadrado}, R. and {Buchlin}, {\'E}. and {Dolla}, L. and {Verbeeck}, C. and {Gissot}, S. and {Teriaca}, L. and {Haberreiter}, M. and {Katsiyannis}, A.~C. and {Rodriguez}, L. and {Kraaikamp}, E. and {Smith}, P.~J. and {Stegen}, K. and {Rochus}, P. and {Halain}, J.~P. and {Jacques}, L. and {Thompson}, W.~T. and {Inhester}, B.},
        title = "{Extreme-UV quiet Sun brightenings observed by the Solar Orbiter/EUI}",
      journal = {\aap},
         year = 2021,
        month = dec,
       volume = {656},
          eid = {L4},
        pages = {L4},
          doi = {10.1051/0004-6361/202140380},
archivePrefix = {arXiv},
       eprint = {2104.03382},
 primaryClass = {astro-ph.SR},
       adsurl = {https://ui.adsabs.harvard.edu/abs/2021A&A...656L...4B}
}

@BOOK{2000mare.book.....P,
       author = {{Priest}, Eric and {Forbes}, Terry},
        title = "{Magnetic Reconnection: MHD Theory and Applications}",
         year = 2000,
          doi = {10.1017/CBO9780511525087},
       adsurl = {https://ui.adsabs.harvard.edu/abs/2000mare.book.....P}
}

@ARTICLE{2017NatCo...815905D,
       author = {{Druett}, Malcolm and {Scullion}, Eamon and {Zharkova}, Valentina and {Matthews}, Sarah and {Zharkov}, Sergei and {Rouppe van der Voort}, Luc},
        title = "{Beam electrons as a source of H{\ensuremath{\alpha}} flare ribbons}",
      journal = {Nature Communications},
         year = 2017,
        month = jun,
       volume = {8},
          eid = {15905},
        pages = {15905},
          doi = {10.1038/ncomms15905},
       adsurl = {https://ui.adsabs.harvard.edu/abs/2017NatCo...815905D}
}

@ARTICLE{1981SoPh...70...25H,
       author = {{Hollweg}, J.~V.},
        title = "{Alfven Waves in the Solar Atmosphere - Part Two - Open and Closed Magnetic Flux Tubes}",
      journal = {\solphys},
         year = 1981,
        month = mar,
       volume = {70},
       number = {1},
        pages = {25-66},
          doi = {10.1007/BF00154391},
       adsurl = {https://ui.adsabs.harvard.edu/abs/1981SoPh...70...25H}
}

@ARTICLE{2015ApJ...805..135T,
       author = {{Takasao}, Shinsuke and {Matsumoto}, Takuma and {Nakamura}, Naoki and {Shibata}, Kazunari},
        title = "{Magnetohydrodynamic Shocks in and above Post-flare Loops: Two-dimensional Simulation and a Simplified Model}",
      journal = {\apj},
         year = 2015,
        month = jun,
       volume = {805},
       number = {2},
          eid = {135},
        pages = {135},
          doi = {10.1088/0004-637X/805/2/135},
archivePrefix = {arXiv},
       eprint = {1504.05700},
 primaryClass = {astro-ph.SR},
       adsurl = {https://ui.adsabs.harvard.edu/abs/2015ApJ...805..135T}
}

@ARTICLE{1983SoPh...84..169F,
       author = {{Forbes}, T.~G. and {Priest}, E.~R.},
        title = "{A Numerical Experiment Relevant to Line-Tied Reconnection in Two-Ribbon Flares}",
      journal = {\solphys},
         year = 1983,
        month = apr,
       volume = {84},
       number = {1-2},
        pages = {169-188},
          doi = {10.1007/BF00157455},
       adsurl = {https://ui.adsabs.harvard.edu/abs/1983SoPh...84..169F}
}

@ARTICLE{2019MNRAS.489.3183C,
       author = {{Cai}, Qiangwei and {Shen}, Chengcai and {Raymond}, John C. and {Mei}, Zhixing and {Warmuth}, Alexander and {Roussev}, Ilia I. and {Lin}, Jun},
        title = "{Investigations of a supra-arcade fan and termination shock above the top of the flare-loop system of the 2017 September 10 event}",
      journal = {\mnras},
         year = 2019,
        month = nov,
       volume = {489},
       number = {3},
        pages = {3183-3199},
          doi = {10.1093/mnras/stz2167},
       adsurl = {https://ui.adsabs.harvard.edu/abs/2019MNRAS.489.3183C}
}

@ARTICLE{2025A&A...702A..88T,
       author = {{Tan}, Song and {Warmuth}, Alexander and {Schuller}, Fr{\'e}d{\'e}ric and {Shen}, Yuandeng and {Ryan}, Daniel F. and {Calchetti}, Daniele and {Hirzberger}, Johann and {Oba}, Takayoshi and {Ulyanov}, Artem and {Valori}, Gherardo},
        title = "{Solar Orbiter reveals persistent magnetic reconnection in medium-scale filament eruptions}",
      journal = {\aap},
         year = 2025,
        month = oct,
       volume = {702},
          eid = {A88},
        pages = {A88},
          doi = {10.1051/0004-6361/202555300},
archivePrefix = {arXiv},
       eprint = {2509.04771},
 primaryClass = {astro-ph.SR},
       adsurl = {https://ui.adsabs.harvard.edu/abs/2025A&A...702A..88T}
}

@ARTICLE{2001EP&S...53..473S,
       author = {{Shibata}, Kazunari and {Tanuma}, Syuniti},
        title = "{Plasmoid-induced-reconnection and fractal reconnection}",
      journal = {Earth, Planets and Space},
         year = 2001,
        month = jun,
       volume = {53},
       number = {6},
        pages = {473-482},
          doi = {10.1186/BF03353258},
archivePrefix = {arXiv},
       eprint = {astro-ph/0101008},
 primaryClass = {astro-ph},
       adsurl = {https://ui.adsabs.harvard.edu/abs/2001EP&S...53..473S}
}

@ARTICLE{2011ApJ...736....3V,
       author = {{van Ballegooijen}, A.~A. and {Asgari-Targhi}, M. and {Cranmer}, S.~R. and {DeLuca}, E.~E.},
        title = "{Heating of the Solar Chromosphere and Corona by Alfv{\'e}n Wave Turbulence}",
      journal = {\apj},
         year = 2011,
        month = jul,
       volume = {736},
       number = {1},
          eid = {3},
        pages = {3},
          doi = {10.1088/0004-637X/736/1/3},
archivePrefix = {arXiv},
       eprint = {1105.0402},
 primaryClass = {astro-ph.SR},
       adsurl = {https://ui.adsabs.harvard.edu/abs/2011ApJ...736....3V}
}

@ARTICLE{2025A&A...701A.294C,
       author = {{Chandra}, Sanghita and {Cameron}, Robert and {Przybylski}, Damien and {Solanki}, Sami K. and {Ondratschek}, Patrick and {Danilovic}, Sanja},
        title = "{Probing chromospheric fine structures with a H{\ensuremath{\alpha}} proxy using MURaM-ChE}",
      journal = {\aap},
         year = 2025,
        month = sep,
       volume = {701},
          eid = {A294},
        pages = {A294},
          doi = {10.1051/0004-6361/202555646},
archivePrefix = {arXiv},
       eprint = {2508.04387},
 primaryClass = {astro-ph.SR},
       adsurl = {https://ui.adsabs.harvard.edu/abs/2025A&A...701A.294C}
}

@article{lysenko2020x,
  title={X-ray and gamma-ray emission from solar flares},
  author={Lysenko, Alexandra L’vovna and Frederiks, Dmitrii Dmitrievich and Fleishman, Gregory D and Aptekar, Rafail L’vovich and Altyntsev, Aleksandr Timofeevich and Golenetskii, Sergei Vladimirovich and Svinkin, Dmitry Sergeevich and Ulanov, Mikhail Vladimirovich and Tsvetkova, Anastasia Evgenievna and Ridnaia, Anna Vladimirovna},
  journal={Physics-Uspekhi},
  volume={63},
  number={8},
  pages={818},
  year={2020},
  publisher={IOP Publishing}
}

@ARTICLE{2023ApJ...947...67R,
       author = {{Ruan}, W. and {Yan}, L. and {Keppens}, R.},
        title = "{Magnetohydrodynamic Turbulence Formation in Solar Flares: 3D Simulation and Synthetic Observations}",
      journal = {\apj},
         year = 2023,
        month = apr,
       volume = {947},
       number = {2},
          eid = {67},
        pages = {67},
          doi = {10.3847/1538-4357/ac9b4e},
archivePrefix = {arXiv},
       eprint = {2210.09856},
 primaryClass = {astro-ph.SR},
       adsurl = {https://ui.adsabs.harvard.edu/abs/2023ApJ...947...67R}
}

@ARTICLE{2023ApJ...955...88Y,
       author = {{Ye}, Jing and {Raymond}, John C. and {Mei}, Zhixing and {Cai}, Qiangwei and {Chen}, Yuhao and {Li}, Yan and {Lin}, Jun},
        title = "{Three-dimensional Simulation of Thermodynamics on Confined Turbulence in a Large-scale CME-flare Current Sheet}",
      journal = {\apj},
         year = 2023,
        month = oct,
       volume = {955},
       number = {2},
          eid = {88},
        pages = {88},
          doi = {10.3847/1538-4357/acf129},
archivePrefix = {arXiv},
       eprint = {2308.09496},
 primaryClass = {astro-ph.SR},
       adsurl = {https://ui.adsabs.harvard.edu/abs/2023ApJ...955...88Y}
}

@ARTICLE{2020A&A...642A...8R,
       author = {{Rochus}, P. and {Auch{\`e}re}, F. and {Berghmans}, D. and {Harra}, L. and {Schmutz}, W. and {Sch{\"u}hle}, U. and {Addison}, P. and {Appourchaux}, T. and {Aznar Cuadrado}, R. and {Baker}, D. and {Barbay}, J. and {Bates}, D. and {BenMoussa}, A. and {Bergmann}, M. and {Beurthe}, C. and {Borgo}, B. and {Bonte}, K. and {Bouzit}, M. and {Bradley}, L. and {B{\"u}chel}, V. and {Buchlin}, E. and {B{\"u}chner}, J. and {Cab{\'e}}, F. and {Cadiergues}, L. and {Chaigneau}, M. and {Chares}, B. and {Choque Cortez}, C. and {Coker}, P. and {Condamin}, M. and {Coumar}, S. and {Curdt}, W. and {Cutler}, J. and {Davies}, D. and {Davison}, G. and {Defise}, J.-M. and {Del Zanna}, G. and {Delmotte}, F. and {Delouille}, V. and {Dolla}, L. and {Dumesnil}, C. and {D{\"u}rig}, F. and {Enge}, R. and {Fran{\c{c}}ois}, S. and {Fourmond}, J.-J. and {Gillis}, J.-M. and {Giordanengo}, B. and {Gissot}, S. and {Green}, L.~M. and {Guerreiro}, N. and {Guilbaud}, A. and {Gyo}, M. and {Haberreiter}, M. and {Hafiz}, A. and {Hailey}, M. and {Halain}, J.-P. and {Hansotte}, J. and {Hecquet}, C. and {Heerlein}, K. and {Hellin}, M.-L. and {Hemsley}, S. and {Hermans}, A. and {Hervier}, V. and {Hochedez}, J.-F. and {Houbrechts}, Y. and {Ihsan}, K. and {Jacques}, L. and {J{\'e}r{\^o}me}, A. and {Jones}, J. and {Kahle}, M. and {Kennedy}, T. and {Klaproth}, M. and {Kolleck}, M. and {Koller}, S. and {Kotsialos}, E. and {Kraaikamp}, E. and {Langer}, P. and {Lawrenson}, A. and {Le Clech'}, J.-C. and {Lenaerts}, C. and {Liebecq}, S. and {Linder}, D. and {Long}, D.~M. and {Mampaey}, B. and {Markiewicz-Innes}, D. and {Marquet}, B. and {Marsch}, E. and {Matthews}, S. and {Mazy}, E. and {Mazzoli}, A. and {Meining}, S. and {Meltchakov}, E. and {Mercier}, R. and {Meyer}, S. and {Monecke}, M. and {Monfort}, F. and {Morinaud}, G. and {Moron}, F. and {Mountney}, L. and {M{\"u}ller}, R. and {Nicula}, B. and {Parenti}, S. and {Peter}, H. and {Pfiffner}, D. and {Philippon}, A. and {Phillips}, I. and {Plesseria}, J.-Y. and {Pylyser}, E. and {Rabecki}, F. and {Ravet-Krill}, M.-F. and {Rebellato}, J. and {Renotte}, E. and {Rodriguez}, L. and {Roose}, S. and {Rosin}, J. and {Rossi}, L. and {Roth}, P. and {Rouesnel}, F. and {Roulliay}, M. and {Rousseau}, A. and {Ruane}, K. and {Scanlan}, J. and {Schlatter}, P. and {Seaton}, D.~B. and {Silliman}, K. and {Smit}, S. and {Smith}, P.~J. and {Solanki}, S.~K. and {Spescha}, M. and {Spencer}, A. and {Stegen}, K. and {Stockman}, Y. and {Szwec}, N. and {Tamiatto}, C. and {Tandy}, J. and {Teriaca}, L. and {Theobald}, C. and {Tychon}, I. and {van Driel-Gesztelyi}, L. and {Verbeeck}, C. and {Vial}, J.-C. and {Werner}, S. and {West}, M.~J. and {Westwood}, D. and {Wiegelmann}, T. and {Willis}, G. and {Winter}, B. and {Zerr}, A. and {Zhang}, X. and {Zhukov}, A.~N.},
        title = "{The Solar Orbiter EUI instrument: The Extreme Ultraviolet Imager}",
      journal = {\aap},
         year = 2020,
        month = oct,
       volume = {642},
          eid = {A8},
        pages = {A8},
          doi = {10.1051/0004-6361/201936663},
       adsurl = {https://ui.adsabs.harvard.edu/abs/2020A&A...642A...8R}
}

@ARTICLE{2014ApJ...791L..13L,
       author = {{Li}, Ting and {Zhang}, Jun},
        title = "{Slipping Magnetic Reconnection Triggering a Solar Eruption of a Triangle-shaped Flag Flux Rope}",
      journal = {\apjl},
         year = 2014,
        month = aug,
       volume = {791},
       number = {1},
          eid = {L13},
        pages = {L13},
          doi = {10.1088/2041-8205/791/1/L13},
archivePrefix = {arXiv},
       eprint = {1407.4180},
 primaryClass = {astro-ph.SR},
       adsurl = {https://ui.adsabs.harvard.edu/abs/2014ApJ...791L..13L}
}

@ARTICLE{2015ApJ...804L...8L,
       author = {{Li}, Ting and {Zhang}, Jun},
        title = "{Quasi-periodic Slipping Magnetic Reconnection During an X-class Solar Flare Observed by the Solar Dynamics Observatory and Interface Region Imaging Spectrograph}",
      journal = {\apjl},
         year = 2015,
        month = may,
       volume = {804},
       number = {1},
          eid = {L8},
        pages = {L8},
          doi = {10.1088/2041-8205/804/1/L8},
archivePrefix = {arXiv},
       eprint = {1504.01111},
 primaryClass = {astro-ph.SR},
       adsurl = {https://ui.adsabs.harvard.edu/abs/2015ApJ...804L...8L}
}

@ARTICLE{1999A&A...351..707T,
       author = {{Titov}, V.~S. and {D{\'e}moulin}, P.},
        title = "{Basic topology of twisted magnetic configurations in solar flares}",
      journal = {\aap},
         year = 1999,
        month = nov,
       volume = {351},
        pages = {707-720},
       adsurl = {https://ui.adsabs.harvard.edu/abs/1999A&A...351..707T}
}

@ARTICLE{1996A&A...308..643D,
       author = {{Demoulin}, P. and {Henoux}, J.~C. and {Priest}, E.~R. and {Mandrini}, C.~H.},
        title = "{Quasi-Separatrix layers in solar flares. I. Method.}",
      journal = {\aap},
         year = 1996,
        month = apr,
       volume = {308},
        pages = {643-655},
       adsurl = {https://ui.adsabs.harvard.edu/abs/1996A&A...308..643D}
}

@ARTICLE{2024ApJ...966...70X,
       author = {{Xing}, Chen and {Aulanier}, Guillaume and {Cheng}, Xin and {Xia}, Chun and {Ding}, Mingde},
        title = "{Unveiling the Initiation Route of Coronal Mass Ejections through Their Slow Rise Phase}",
      journal = {\apj},
         year = 2024,
        month = may,
       volume = {966},
       number = {1},
          eid = {70},
        pages = {70},
          doi = {10.3847/1538-4357/ad2ea9},
archivePrefix = {arXiv},
       eprint = {2402.16679},
 primaryClass = {astro-ph.SR},
       adsurl = {https://ui.adsabs.harvard.edu/abs/2024ApJ...966...70X}
}

@ARTICLE{2025ApJ...986...37X,
       author = {{Xing}, Chen and {Cheng}, Xin and {Aulanier}, Guillaume and {Ding}, Mingde},
        title = "{Initiation Route of Coronal Mass Ejections. II. The Role of Filament Mass}",
      journal = {\apj},
         year = 2025,
        month = jun,
       volume = {986},
       number = {1},
          eid = {37},
        pages = {37},
          doi = {10.3847/1538-4357/adceb5},
archivePrefix = {arXiv},
       eprint = {2504.14876},
 primaryClass = {astro-ph.SR},
       adsurl = {https://ui.adsabs.harvard.edu/abs/2025ApJ...986...37X}
}

@ARTICLE{2002SoPh..210..307F,
       author = {{Fletcher}, L. and {Hudson}, H.~S.},
        title = "{Spectral and Spatial Variations of Flare Hard X-ray Footpoints}",
      journal = {\solphys},
         year = 2002,
        month = nov,
       volume = {210},
       number = {1},
        pages = {307-321},
          doi = {10.1023/A:1022479610710},
       adsurl = {https://ui.adsabs.harvard.edu/abs/2002SoPh..210..307F}
}

@ARTICLE{2007ApJ...660..863T,
       author = {{Titov}, V.~S.},
        title = "{Generalized Squashing Factors for Covariant Description of Magnetic Connectivity in the Solar Corona}",
      journal = {\apj},
         year = 2007,
        month = may,
       volume = {660},
       number = {1},
        pages = {863-873},
          doi = {10.1086/512671},
archivePrefix = {arXiv},
       eprint = {astro-ph/0703671},
 primaryClass = {astro-ph},
       adsurl = {https://ui.adsabs.harvard.edu/abs/2007ApJ...660..863T}
}

@ARTICLE{2017JPlPh..83a5301J,
       author = {{Janvier}, Miho},
        title = "{Three-dimensional magnetic reconnection and its application to solar flares}",
      journal = {Journal of Plasma Physics},
         year = 2017,
        month = feb,
       volume = {83},
       number = {1},
          eid = {535830101},
        pages = {535830101},
          doi = {10.1017/S0022377817000034},
archivePrefix = {arXiv},
       eprint = {1612.06513},
 primaryClass = {astro-ph.SR},
       adsurl = {https://ui.adsabs.harvard.edu/abs/2017JPlPh..83a5301J}
}

@ARTICLE{2009ApJ...700..559M,
       author = {{Masson}, S. and {Pariat}, E. and {Aulanier}, G. and {Schrijver}, C.~J.},
        title = "{The Nature of Flare Ribbons in Coronal Null-Point Topology}",
      journal = {\apj},
         year = 2009,
        month = jul,
       volume = {700},
       number = {1},
        pages = {559-578},
          doi = {10.1088/0004-637X/700/1/559},
       adsurl = {https://ui.adsabs.harvard.edu/abs/2009ApJ...700..559M}
}

@ARTICLE{Lemmen2012,
       author = {{Lemen}, James R. and {Title}, Alan M. and {Akin}, David J. and {Boerner}, Paul F. and {Chou}, Catherine and {Drake}, Jerry F. and {Duncan}, Dexter W. and {Edwards}, Christopher G. and {Friedlaender}, Frank M. and {Heyman}, Gary F. and {Hurlburt}, Neal E. and {Katz}, Noah L. and {Kushner}, Gary D. and {Levay}, Michael and {Lindgren}, Russell W. and {Mathur}, Dnyanesh P. and {McFeaters}, Edward L. and {Mitchell}, Sarah and {Rehse}, Roger A. and {Schrijver}, Carolus J. and {Springer}, Larry A. and {Stern}, Robert A. and {Tarbell}, Theodore D. and {Wuelser}, Jean-Pierre and {Wolfson}, C. Jacob and {Yanari}, Carl and {Bookbinder}, Jay A. and {Cheimets}, Peter N. and {Caldwell}, David and {Deluca}, Edward E. and {Gates}, Richard and {Golub}, Leon and {Park}, Sang and {Podgorski}, William A. and {Bush}, Rock I. and {Scherrer}, Philip H. and {Gummin}, Mark A. and {Smith}, Peter and {Auker}, Gary and {Jerram}, Paul and {Pool}, Peter and {Soufli}, Regina and {Windt}, David L. and {Beardsley}, Sarah and {Clapp}, Matthew and {Lang}, James and {Waltham}, Nicholas},
        title = "{The Atmospheric Imaging Assembly (AIA) on the Solar Dynamics Observatory (SDO)}",
      journal = {\solphys},
         year = 2012,
        month = jan,
       volume = {275},
       number = {1-2},
        pages = {17-40},
          doi = {10.1007/s11207-011-9776-8},
       adsurl = {https://ui.adsabs.harvard.edu/abs/2012SoPh..275...17L}
}

@ARTICLE{Pesnell2012,
       author = {{Pesnell}, W. Dean and {Thompson}, B.~J. and {Chamberlin}, P.~C.},
        title = "{The Solar Dynamics Observatory (SDO)}",
      journal = {\solphys},
         year = 2012,
        month = jan,
       volume = {275},
       number = {1-2},
        pages = {3-15},
          doi = {10.1007/s11207-011-9841-3},
       adsurl = {https://ui.adsabs.harvard.edu/abs/2012SoPh..275....3P}
}

@ARTICLE{Pietrow2024,
       author = {{Pietrow}, A.~G.~M. and {Cretignier}, M. and {Druett}, M.~K. and {Alvarado-G{\'o}mez}, J.~D. and {Hofmeister}, S.~J. and {Verma}, M. and {Kamlah}, R. and {Baratella}, M. and {Amazo-G{\'o}mez}, E.~M. and {Kontogiannis}, I. and {Dineva}, E. and {Warmuth}, A. and {Denker}, C. and {Poppenhaeger}, K. and {Andriienko}, O. and {Dumusque}, X. and {L{\"o}fdahl}, M.~G.},
        title = "{A comparative study of two X2.2 and X9.3 solar flares observed with HARPS-N. Reconciling Sun-as-a-star spectroscopy and high-spatial resolution solar observations in the context of the solar-stellar connection}",
      journal = {\aap},
         year = 2024,
        month = feb,
       volume = {682},
          eid = {A46},
        pages = {A46},
          doi = {10.1051/0004-6361/202347895},
archivePrefix = {arXiv},
       eprint = {2309.03373},
 primaryClass = {astro-ph.SR},
       adsurl = {https://ui.adsabs.harvard.edu/abs/2024A&A...682A..46P}
}

@ARTICLE{2014DePontieuIRIS,
       author = {{De Pontieu}, B. and {Title}, A.~M. and {Lemen}, J.~R. and {Kushner}, G.~D. and {Akin}, D.~J. and {Allard}, B. and {Berger}, T. and {Boerner}, P. and {Cheung}, M. and {Chou}, C. and {Drake}, J.~F. and {Duncan}, D.~W. and {Freeland}, S. and {Heyman}, G.~F. and {Hoffman}, C. and {Hurlburt}, N.~E. and {Lindgren}, R.~W. and {Mathur}, D. and {Rehse}, R. and {Sabolish}, D. and {Seguin}, R. and {Schrijver}, C.~J. and {Tarbell}, T.~D. and {W{\"u}lser}, J. -P. and {Wolfson}, C.~J. and {Yanari}, C. and {Mudge}, J. and {Nguyen-Phuc}, N. and {Timmons}, R. and {van Bezooijen}, R. and {Weingrod}, I. and {Brookner}, R. and {Butcher}, G. and {Dougherty}, B. and {Eder}, J. and {Knagenhjelm}, V. and {Larsen}, S. and {Mansir}, D. and {Phan}, L. and {Boyle}, P. and {Cheimets}, P.~N. and {DeLuca}, E.~E. and {Golub}, L. and {Gates}, R. and {Hertz}, E. and {McKillop}, S. and {Park}, S. and {Perry}, T. and {Podgorski}, W.~A. and {Reeves}, K. and {Saar}, S. and {Testa}, P. and {Tian}, H. and {Weber}, M. and {Dunn}, C. and {Eccles}, S. and {Jaeggli}, S.~A. and {Kankelborg}, C.~C. and {Mashburn}, K. and {Pust}, N. and {Springer}, L. and {Carvalho}, R. and {Kleint}, L. and {Marmie}, J. and {Mazmanian}, E. and {Pereira}, T.~M.~D. and {Sawyer}, S. and {Strong}, J. and {Worden}, S.~P. and {Carlsson}, M. and {Hansteen}, V.~H. and {Leenaarts}, J. and {Wiesmann}, M. and {Aloise}, J. and {Chu}, K. -C. and {Bush}, R.~I. and {Scherrer}, P.~H. and {Brekke}, P. and {Martinez-Sykora}, J. and {Lites}, B.~W. and {McIntosh}, S.~W. and {Uitenbroek}, H. and {Okamoto}, T.~J. and {Gummin}, M.~A. and {Auker}, G. and {Jerram}, P. and {Pool}, P. and {Waltham}, N.},
        title = "{The Interface Region Imaging Spectrograph (IRIS)}",
      journal = {\solphys},
         year = 2014,
        month = jul,
       volume = {289},
       number = {7},
        pages = {2733-2779},
          doi = {10.1007/s11207-014-0485-y},
archivePrefix = {arXiv},
       eprint = {1401.2491},
 primaryClass = {astro-ph.SR},
       adsurl = {https://ui.adsabs.harvard.edu/abs/2014SoPh..289.2733D}
}

@ARTICLE{2014ApJ...788L..18S,
       author = {{Sharykin}, I.~N. and {Kosovichev}, A.~G.},
        title = "{Fine Structure of Flare Ribbons and Evolution of Electric Currents}",
      journal = {\apjl},
         year = 2014,
        month = jun,
       volume = {788},
       number = {1},
          eid = {L18},
        pages = {L18},
          doi = {10.1088/2041-8205/788/1/L18},
archivePrefix = {arXiv},
       eprint = {1404.5104},
 primaryClass = {astro-ph.SR},
       adsurl = {https://ui.adsabs.harvard.edu/abs/2014ApJ...788L..18S}
}

@ARTICLE{2022FrASS...940945L,
       author = {{L{\"o}rin{\v{c}}{\'\i}k}, Juraj and {Polito}, Vanessa and {De Pontieu}, Bart and {Yu}, Sijie and {Freij}, Nabil},
        title = "{Rapid variations of Si IV spectra in a flare observed by interface region imaging spectrograph at a sub-second cadence}",
      journal = {Frontiers in Astronomy and Space Sciences},
         year = 2022,
        month = nov,
       volume = {9},
          eid = {334},
        pages = {334},
          doi = {10.3389/fspas.2022.1040945},
archivePrefix = {arXiv},
       eprint = {2210.12205},
 primaryClass = {astro-ph.SR},
       adsurl = {https://ui.adsabs.harvard.edu/abs/2022FrASS...940945L}
}

@ARTICLE{2025A&A...693A...8T,
       author = {{Thoen Faber}, Jonas and {Joshi}, Reetika and {Rouppe van der Voort}, Luc and {Wedemeyer}, Sven and {Fletcher}, Lyndsay and {Aulanier}, Guillaume and {N{\'o}brega-Siverio}, Daniel},
        title = "{High-resolution observational analysis of flare ribbon fine structures}",
      journal = {\aap},
         year = 2025,
        month = jan,
       volume = {693},
          eid = {A8},
        pages = {A8},
          doi = {10.1051/0004-6361/202452370},
archivePrefix = {arXiv},
       eprint = {2411.18233},
 primaryClass = {astro-ph.SR},
       adsurl = {https://ui.adsabs.harvard.edu/abs/2025A&A...693A...8T}
}

@ARTICLE{2025ApJ...982L...9Z,
       author = {{Zhang}, Yining and {Li}, Ting and {Hou}, Yijun and {Duan}, Xuchun and {Sun}, Zheng and {Zhou}, Guiping},
        title = "{Quasiperiodic Super-Alfv{\'e}nic Slippage along Flare Ribbons Observed by the Interface Region Imaging Spectrograph}",
      journal = {\apjl},
         year = 2025,
        month = mar,
       volume = {982},
       number = {1},
          eid = {L9},
        pages = {L9},
          doi = {10.3847/2041-8213/adbb6e},
archivePrefix = {arXiv},
       eprint = {2502.16579},
 primaryClass = {astro-ph.SR},
       adsurl = {https://ui.adsabs.harvard.edu/abs/2025ApJ...982L...9Z}
}

@ARTICLE{2013ApJ...766..127Y,
       author = {{Young}, P.~R. and {Doschek}, G.~A. and {Warren}, H.~P. and {Hara}, H.},
        title = "{Properties of a Solar Flare Kernel Observed by Hinode and SDO}",
      journal = {\apj},
         year = 2013,
        month = apr,
       volume = {766},
       number = {2},
          eid = {127},
        pages = {127},
          doi = {10.1088/0004-637X/766/2/127},
archivePrefix = {arXiv},
       eprint = {1212.4388},
 primaryClass = {astro-ph.SR},
       adsurl = {https://ui.adsabs.harvard.edu/abs/2013ApJ...766..127Y}
}

@ARTICLE{2016ApJ...823...62Z,
       author = {{Zhao}, Jie and {Gilchrist}, Stuart A. and {Aulanier}, Guillaume and {Schmieder}, Brigitte and {Pariat}, Etienne and {Li}, Hui},
        title = "{Hooked Flare Ribbons and Flux-rope-related QSL Footprints}",
      journal = {\apj},
         year = 2016,
        month = may,
       volume = {823},
       number = {1},
          eid = {62},
        pages = {62},
          doi = {10.3847/0004-637X/823/1/62},
archivePrefix = {arXiv},
       eprint = {1603.07563},
 primaryClass = {astro-ph.SR},
       adsurl = {https://ui.adsabs.harvard.edu/abs/2016ApJ...823...62Z}
}

@ARTICLE{2025ApJ...990...45S,
       author = {{Sun}, Zheng and {Li}, Ting and {Bian}, Xinkai and {Hou}, Yijun and {Kontogiannis}, Ioannis and {Wu}, Ziqi},
        title = "{Current Helicity in Response to Coronal Mass Ejections}",
      journal = {\apj},
         year = 2025,
        month = sep,
       volume = {990},
       number = {1},
          eid = {45},
        pages = {45},
          doi = {10.3847/1538-4357/adf18b},
archivePrefix = {arXiv},
       eprint = {2507.11790},
 primaryClass = {astro-ph.SR},
       adsurl = {https://ui.adsabs.harvard.edu/abs/2025ApJ...990...45S}
}

@ARTICLE{2015ApJ...811..139T,
       author = {{Tian}, Hui and {Young}, Peter R. and {Reeves}, Katharine K. and {Chen}, Bin and {Liu}, Wei and {McKillop}, Sean},
        title = "{Temporal Evolution of Chromospheric Evaporation: Case Studies of the M1.1 Flare on 2014 September 6 and X1.6 Flare on 2014 September 10}",
      journal = {\apj},
         year = 2015,
        month = oct,
       volume = {811},
       number = {2},
          eid = {139},
        pages = {139},
          doi = {10.1088/0004-637X/811/2/139},
archivePrefix = {arXiv},
       eprint = {1505.02736},
 primaryClass = {astro-ph.SR},
       adsurl = {https://ui.adsabs.harvard.edu/abs/2015ApJ...811..139T}
}

@ARTICLE{2014ApJ...797L..14T,
       author = {{Tian}, Hui and {Li}, Gang and {Reeves}, Katharine K. and {Raymond}, John C. and {Guo}, Fan and {Liu}, Wei and {Chen}, Bin and {Murphy}, Nicholas A.},
        title = "{Imaging and Spectroscopic Observations of Magnetic Reconnection and Chromospheric Evaporation in a Solar Flare}",
      journal = {\apjl},
         year = 2014,
        month = dec,
       volume = {797},
       number = {2},
          eid = {L14},
        pages = {L14},
          doi = {10.1088/2041-8205/797/2/L14},
archivePrefix = {arXiv},
       eprint = {1411.2301},
 primaryClass = {astro-ph.SR},
       adsurl = {https://ui.adsabs.harvard.edu/abs/2014ApJ...797L..14T}
}

@ARTICLE{2019ApJ...879...30L,
       author = {{Li}, Y. and {Ding}, M.~D. and {Hong}, J. and {Li}, H. and {Gan}, W.~Q.},
        title = "{Different Signatures of Chromospheric Evaporation in Two Solar Flares Observed with IRIS}",
      journal = {\apj},
         year = 2019,
        month = jul,
       volume = {879},
       number = {1},
          eid = {30},
        pages = {30},
          doi = {10.3847/1538-4357/ab245a},
archivePrefix = {arXiv},
       eprint = {1905.09709},
 primaryClass = {astro-ph.SR},
       adsurl = {https://ui.adsabs.harvard.edu/abs/2019ApJ...879...30L}
}

@ARTICLE{2016ApJ...823L..13L,
       author = {{Li}, Y. and {Qiu}, J. and {Longcope}, D.~W. and {Ding}, M.~D. and {Yang}, K.},
        title = "{Observations of an X-shaped Ribbon Flare in the Sun and Its Three-dimensional Magnetic Reconnection}",
      journal = {\apjl},
         year = 2016,
        month = may,
       volume = {823},
       number = {1},
          eid = {L13},
        pages = {L13},
          doi = {10.3847/2041-8205/823/1/L13},
archivePrefix = {arXiv},
       eprint = {1605.01833},
 primaryClass = {astro-ph.SR},
       adsurl = {https://ui.adsabs.harvard.edu/abs/2016ApJ...823L..13L}
}

@ARTICLE{2017ApJ...845L..18D,
       author = {{De Pontieu}, B. and {De Moortel}, I. and {Martinez-Sykora}, J. and {McIntosh}, S.~W.},
        title = "{Observations and Numerical Models of Solar Coronal Heating Associated with Spicules}",
      journal = {\apjl},
         year = 2017,
        month = aug,
       volume = {845},
       number = {2},
          eid = {L18},
        pages = {L18},
          doi = {10.3847/2041-8213/aa7fb4},
archivePrefix = {arXiv},
       eprint = {1710.06790},
 primaryClass = {astro-ph.SR},
       adsurl = {https://ui.adsabs.harvard.edu/abs/2017ApJ...845L..18D}
}

\appendix
\section{An example of the parameter measurement} \label{sec:tracker}
To determine the projected velocities and lifetimes of the downflows, we manually tracked the front of each feature on a frame-by-frame basis using the open-source software \texttt{Tracker}. The software serves as an interactive tool to facilitate the identification of feature trajectories, while the underlying method remains a manual tracking procedure similar to those adopted in previous studies (e.g., \citealt{2022MNRAS.516.3120T,2025arXiv250701169S,2025A&A...702A.189T}).

Since \texttt{Tracker} operates on video files, the original dataset is first converted into a video and then imported into the software. An $x$--$y$ coordinate system and a spatial scale are added. In the present analysis, the absolute choice of the coordinate origin and axis orientation is not important, since we only require the relative displacement of the tracked features. After these steps, the front of each downflow feature is manually identified and tracked throughout its evolution. Figure~A.1 shows a representative example of the tracking procedure adopted in this work.
Using the manually selected front positions in successive frames, we calculate the projected speed of the downflow. Since no clear acceleration or deceleration is found in most cases, we simply average the measured speeds among all the frames and use this value as the characteristic speed of the downflow. The lifetime is determined from the total duration over which the downflow can be identified.
To estimate the downflow length, we use the last frame of the observed downflow. The spine of the downflow is manually approximated with a straight line. We then define the edge of the downflow as the position where the intensity decreases to 30\% of the brightest front intensity along the spine direction. The resulting distance is adopted as the length of the downflow.
\begin{figure}
   \centering
     \includegraphics[width=1\columnwidth]{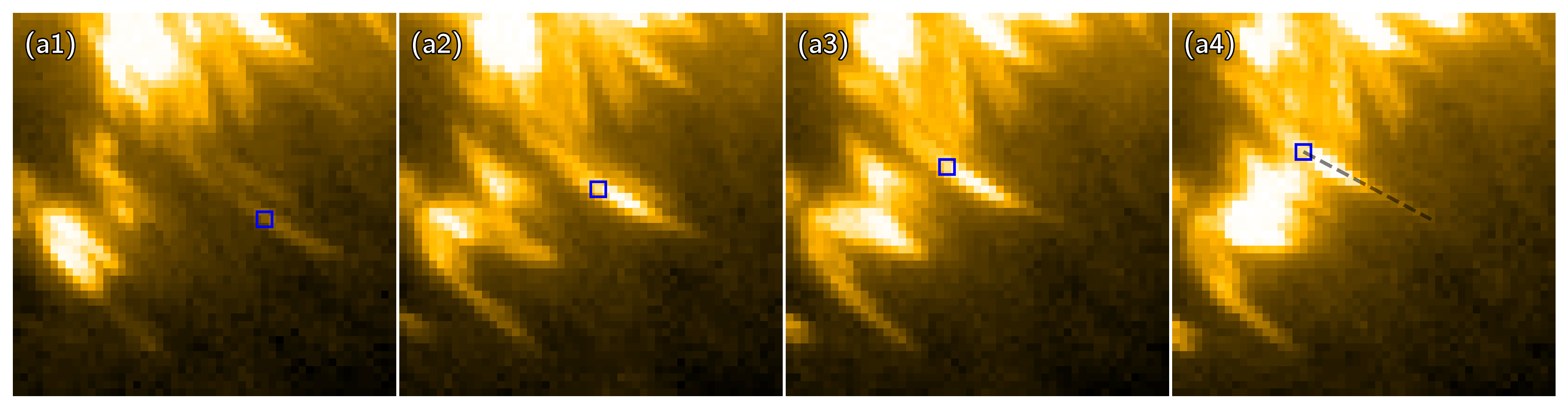}
     \caption{Example of the frame-by-frame tracking procedure. The dark blue squares indicate the manually identified front of the downflow in successive frames, while the black dashed line in the last panel indicates the manually measured length of the downflow.
   \label{fig:Fig.appendix}}
 \end{figure}
 
\end{document}